\documentclass[onecolumn,useAMS,usenatbib,usegraphicx]{mn2e}

\usepackage{amssymb}
\usepackage{tabularx}

\def\l{{\ell}}
\def\lm{{\l m}}
\def\summ{\sum_{m=-\ell}^{\ell}}
\def\suml{\sum_{\ell=0}^{\infty}}
\def\alm{a_{\lm}}
\def\plm{\Phi_{\lm}}
\def\ylm{Y_{\lm}}

\def\Tp{T_{\theta,\phi}}
\def\healpix{H{\sc ealpix }}

\newcommand{\cardiff}{{School of Physics \& Astronomy, Cardiff University,
    5 The Parade, Cardiff, CF24 3AA, Wales, United Kingdom}}

\newcommand{\capetowna}{{Department of Mathematics and Applied Mathematics, University of Cape Town, Rondebosch 7701, Cape Town, Republic of South Africa}}

\newcommand{\capetownb}{{Astrophysics, Cosmology and Gravity Centre (ACGC), University of Cape Town, Rondebosch 7701, Cape Town, Republic of South Africa}}

\title[Statistical Characterization of Temperature Patterns in Anisotropic Cosmologies]
  {Statistical characterization of temperature patterns in anisotropic cosmologies}
\author[Sung, Short \& Coles]
  {Rockhee Sung$^{1,2,3}$, Jo Short$^{1}$, and Peter Coles$^{1}$ \\  \\
$^1$  \cardiff \\
$^2$  \capetowna \\
$^3$  \capetownb \\}

\date{Accepted 2010 ???? ???; Received 2010 ???? ???}

\begin{document}

\maketitle

\begin{abstract}
We consider the issue of characterizing the coherent large-scale patterns from CMB temperature maps in globally anisotropic cosmologies. The methods we investigate are reasonably general; the particular models we test them on are the homogeneous but anisotropic relativistic cosmologies described by the Bianchi classification. Although the temperature variations produced in these models are not stochastic, they give rise to a ``non--Gaussian'' distribution of temperature fluctuations over the sky that is a partial diagnostic of the model. We explore two methods for quantifying non--Gaussian and/or non-stationary fluctuation fields in order to see how they respond to the Bianchi models. We first investigate the behavior of phase correlations between the spherical harmonic modes of the maps. Then we examine the behavior of the multipole vectors of the temperature distribution which, though defined in harmonic space, can indicate the presence of a preferred direction in real space, i.e. on the 2-sphere. These methods give extremely clear signals of the presence of anisotropy when applied to the models we discuss, suggesting that they have some promise as diagnostics of the presence of global asymmetry in the Universe.
\end{abstract}


\section{Introduction}
\label{secIntro} Observations of the Cosmic Microwave Background
(CMB) provide some of the most compelling support for the currently
favored $\Lambda$CDM, or \emph{concordance}, cosmological model. The
concordance framework predicts that the CMB should posses
temperature fluctuations which are both statistically isotropic
(i.e. stationary over the celestial sphere) and Gaussian
\citep{Guth1982,Starobinskij1982,Bardeen1983}. Measurements by the
Wilkinson Microwave Anisotropy Probe (WMAP)
\citep{Bennett2003,Hinshaw2009} have undergone extensive statistical
analysis, much of which has confirmed the concordance model but with
some indications of departures that may be significant; see for
example \cite{Yadav2008}. More specifically, there is some evidence
for hemispherical power asymmetry
\citep{Eriksen2004a,Park2004,Eriksen2007,Hoftuft2009,Hansen2009} and
also a Cold Spot has been identified \citep{Vielva2004,Cruz2005}. In
other words there is some evidence of an anisotropic universe, i.e.
one in which the background cosmology may not be described by the
standard Friedman-Robertson-Walker (FRW) metric. Of course the
background cosmology for a non-isotropic universe may still be
described by the FRW metric, but this would require a non-standard
topology which we do not consider in this analysis.

The Bianchi classification provides a complete characterization of
all the known homogeneous but anisotropic exact solutions to General
Relativity. The classification was first proposed by Bianchi and
later applied to General Relativity \citep{Ellis1969}. Initial
studies used the lack of large-scale asymmetry in the CMB
temperature to put strong constraints on the possible Bianchi models
\citep{Barrow1985, Bunn1996, Kogut1997}. However, simulations of the
CMB from Bianchi universes not only show a preferred direction, but
models with negative spatial curvature (such as the types V and
VII$_h$) can produce localized features \citep{Barrow1985}. So more
recently attention has shifted to reproducing a Cold Spot such as
that claimed to exist in the WMAP data. Initially, Type VII$_h$ was
the favored model to best reproduce the anomaly \citep{Jaffe2005,
Jaffe2006a,Jaffe2006b}, and this has subsequently been investigated
quite thoroughly \citep{McEwen1,McEwen2,Pontz1,Pontz2,Sung2010},
although more recent work has also looked at the Bianchi Type V
which also produces localized features \citep{Sung2009}.

\begin{figure}
\begin{centering}
  \includegraphics[width=58mm]{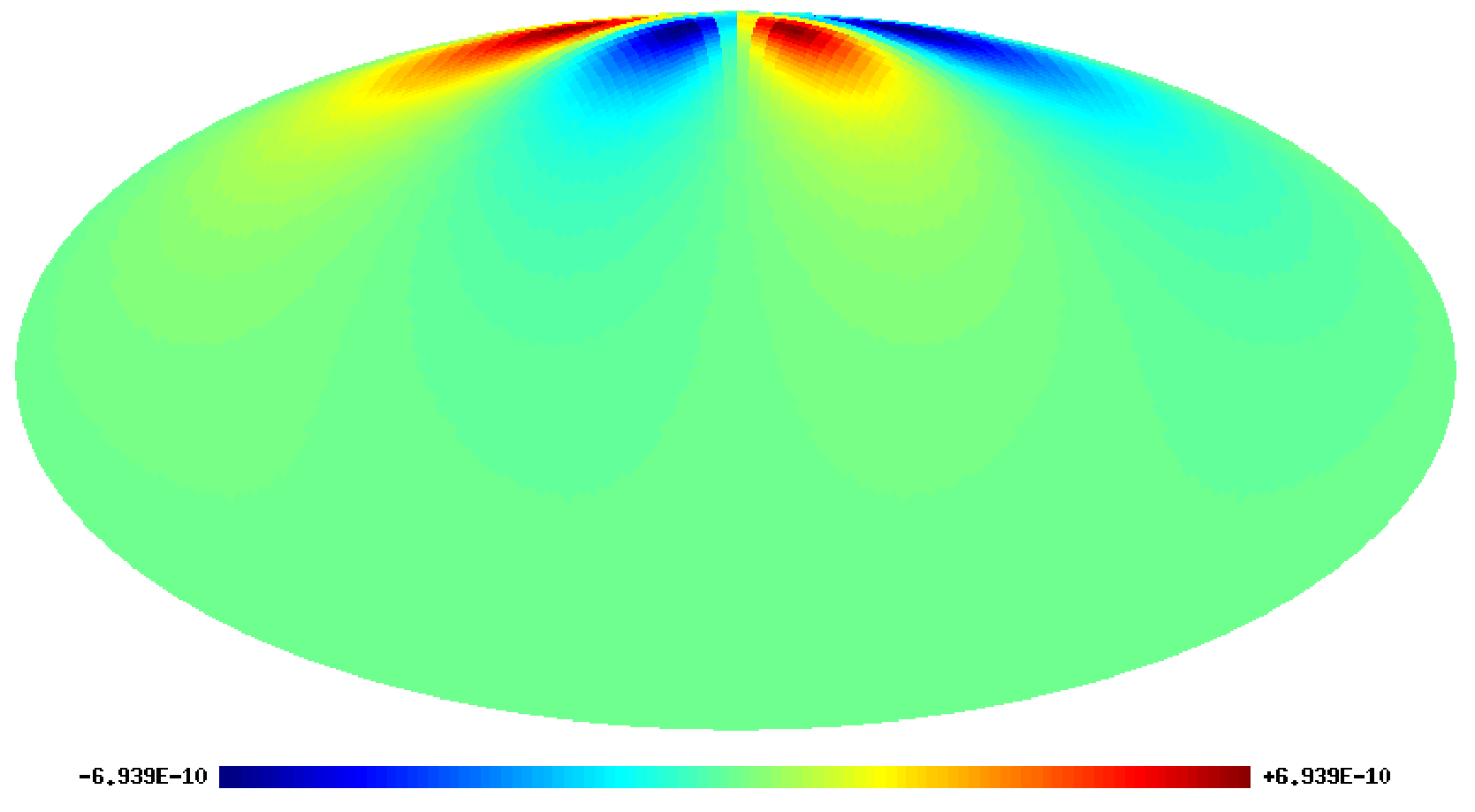}
  \includegraphics[width=58mm]{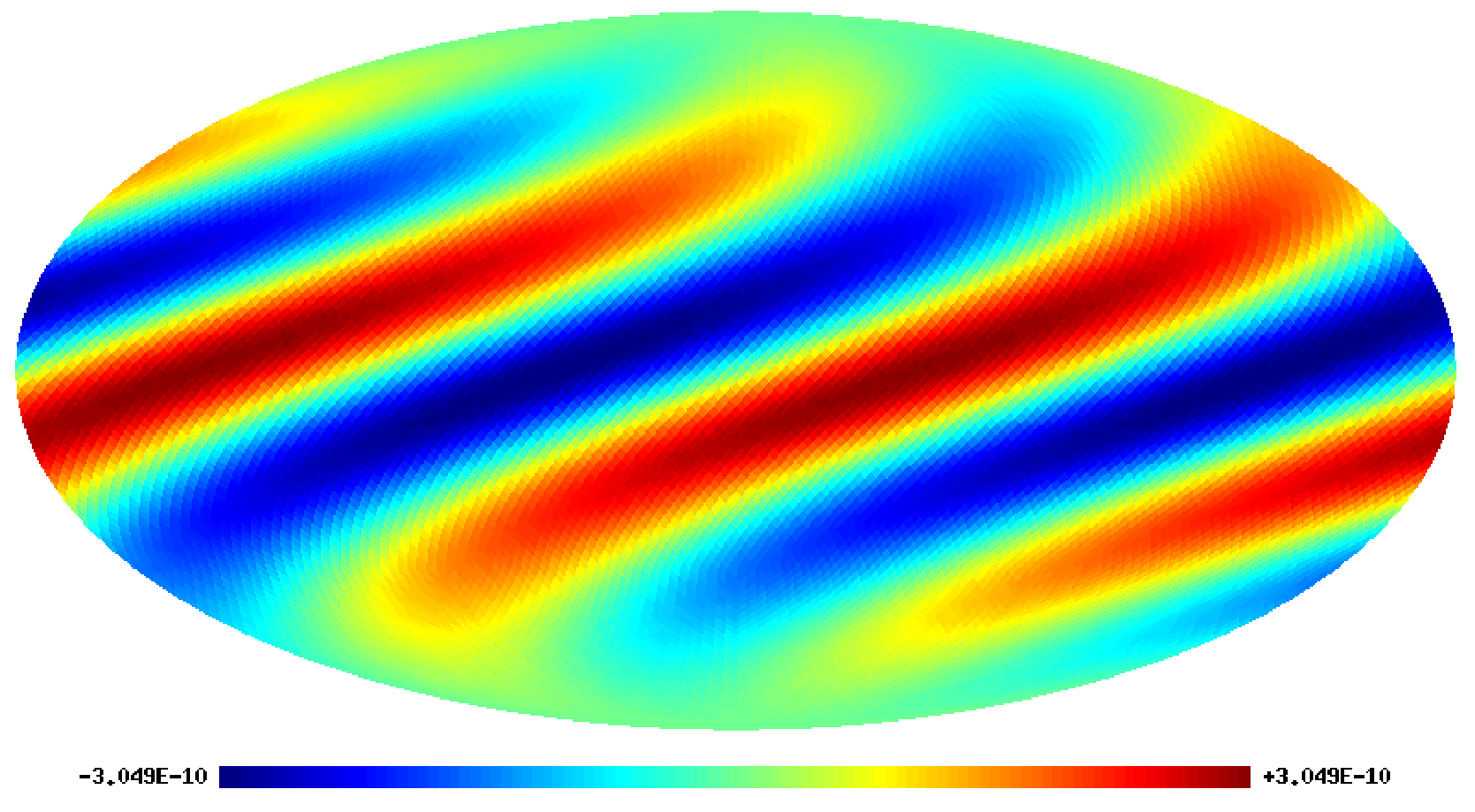}
  \includegraphics[width=58mm]{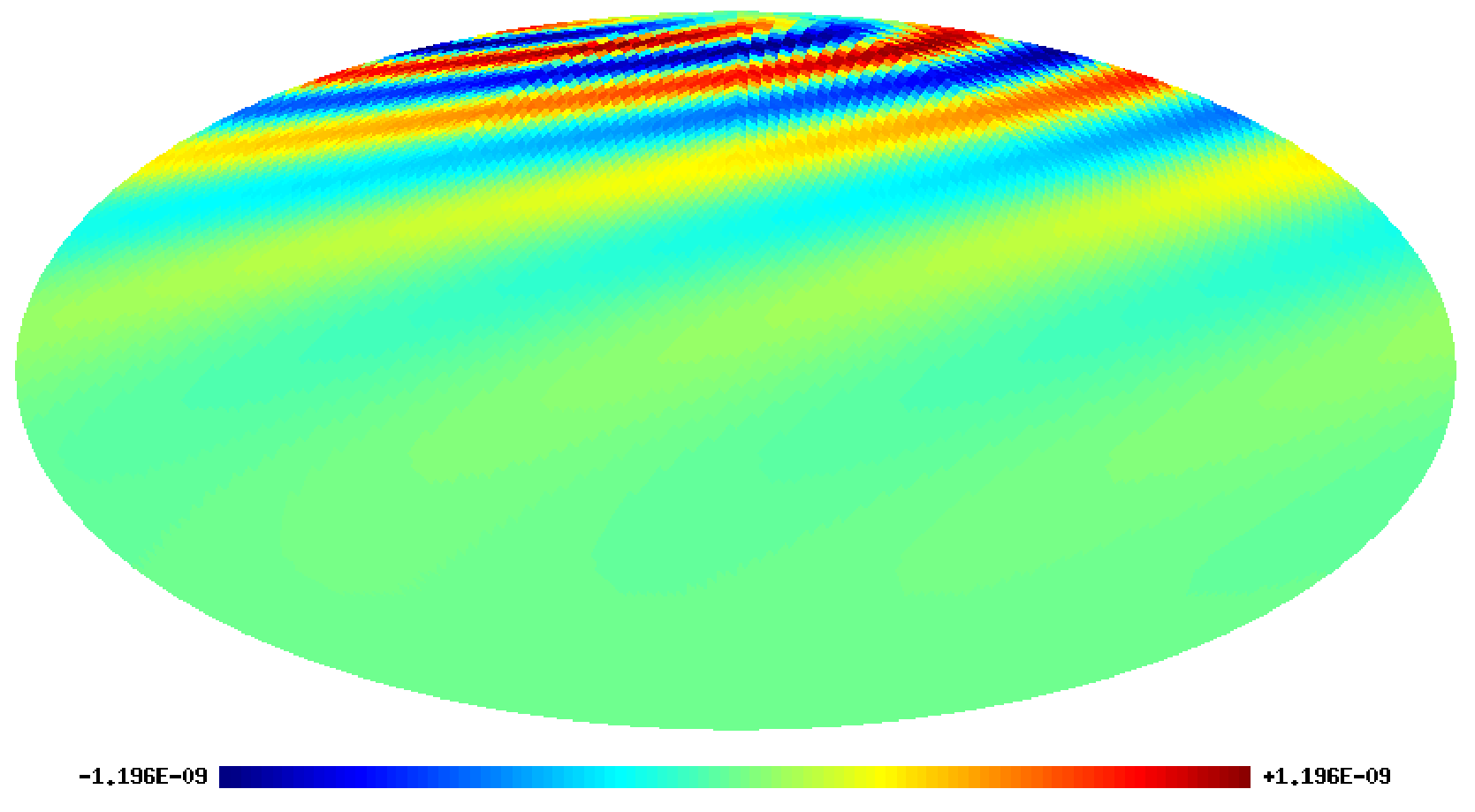}.
  \caption{Simulated maps of the the CMB temperature, at redshift z = 0, using Bianchi type cosmologies. From left to right the Bianchi types are: V, VII$_0$ and VII$_h$. The colour scale is marked in milliKelvin. All the maps started as a quadruple at z = 500. The Bianchi V map shows a focused feature, the Bianchi VII$_0$ map has a twisted feature and the Bianchi VII$_h$ map has both focusing and twisting in the resulting temperature pattern.}
  \label{figBT}
\end{centering}
\end{figure}

The most interesting range of anisotropic structures is produced in
Bianchi Types VII$_h$, VII$_0$ and V. These different Bianchi types
have the effect of focusing and/or twisting the initial quadrupole
over time (see Figure \ref{figBT}). In this paper we study the
behavior of these Bianchi models so as to identify characteristics
of the radiation fields they produce and develop methods that can be
used to identify more general forms of anisotropy. Understanding the
characteristics identified in these particular cases will hopefully
help us find better and more systematic ways of constraining the
level of global symmetry present in the real Universe. Note we
consider just characteristics observable in the CMB temperature; we
shall return to a study of the polarization radiation component in
later work.

We consider two statistical measures of anisotropy in some detail in
this paper. Neither of these is entirely new and both have
previously been applied to observed CMB maps. However, the general
philosophy behind previous applications of these methods has been
simply to look for departures from the (composite) null hypothesis
of statistical isotropy and Gaussianity (or more recently they have
been developed to look at universes with multiply-connected
topologies \citep{Bielewicz2009}). In other words, they have been
used to construct hypothesis tests with the concordance cosmology
but their performance has not hitherto been evaluated on models with
built-in anisotropy.

For example, if the concordance model is correct, the {\em phases}
of the spherical harmonic coefficients of the CMB should be
independently random and uniformly distributed. Recent studies have
suggested some deviation from this
\citep{Coles2004,sc2005,dc2005,Chiang2007,ccno7} but it is not clear
whether they indicate global anisotropy or departures from
Gaussianity, let alone whether these are of cosmic or instrumental
origin. Here we examine the use of phase correlations in quantifying
the temperature patterns generated in models with known levels of
global inhomogeneity.

Multipole vectors were first introduced over a  century ago
\citep{Maxwell1891}. There have since been attempts to understand
the multipole vectors in order to explain the CMB anomalies reported
at large angular scales \citep{Katz2004, Schwarz2004, Copi2004,
Land2005a, Land2005b, Land2005c, Land2005d, Land2005e, Copi2006,
Copi2007} since is not clear how to quantify and verify such
properties from the CMB anomalies in spherical harmonics. They have
been used in a number of studies to show anomalies, such as
alignments of multiples \citep{Abramo2006} in a similar plane to the
axis of evil \citep{Land2005d,Land2007}. Our aim here is to examine
the behavior of the multipole vectors in cases where the form of
anisotropy is known {\em priori} in order to assess their potential
to act as more general descriptors.

Two points are worth making before we continue. First,  any
realistic cosmology (whether of FRW or Bianchi type) will possess
random fluctuations on top of a smooth background. If these
fluctuations are stationary Gaussian then they will add correlated
``noise'' to any signal arising from the background model and will
thus hamper the performance of any statistical analysis method,
especially at smaller angular scales. This Gaussian ``noise'' (which is
equivalent to stationary Gaussian fluctuations, and not to be
confused with instrumental noise) is completely characterized by
second-order statistical quantities (i.e. the power spectrum in
harmonic space or the autocorrelation function in pixel space). The
statistical descriptors we explore are {\em independent} of the
power-spectrum, so adding Gaussian noise will not produce any
systematic response in them. 

We also restrict ourselves to looking at just the large-scale features
because the patterns in the temperature maps resulting from the Bianchi models
is over large scales. Therefore, by looking at large scales only, there
is more chance of detecting the anisotropy. However, it goes without saying we are
not claiming that these Bianchi models are in themselves complete
alternatives to the concordance cosmology. Rather we think of them
as representing possible perturbation modes of the FRW background.

The layout of this article is as follows. In Section \ref{secPixel}
we look at pixel  distributions of the CMB maps to show how the
statistical anisotropy present in these models produces a form of
non-Gaussianity in the pixel distribution over the celestial sphere.
We then introduce phase correlations in Section \ref{secPhase} to
provide characterization of the anisotropy displayed by the models.
In Section \ref{secMultipole} we look at the behavior of the
multipole vectors as characteristics of the anisotropy of the same
maps. Finally, Section \ref{secConclu} summarizes the conclusions.

\section{Statistical Descriptors of non-Gaussianity}
\label{secPixel} \cite{Sung2009} discussed how localized features in
the CMB temperature pattern,  perhaps similar to the Cold Spot
observed in the WMAP data \citep{Vielva2004,Cruz2005}, can be
generated in models with negative spatial curvature, i.e. Bianchi
types V and VII$_{h}$. In the standard cosmological framework the
temperature fluctuations are described by a Gaussian random process
over the sky, so a feature like the cold spot corresponds to an
extreme event in the tail of the distribution of fluctuations. In a
Bianchi model, however, it is not stochastic at all but produced
{\em coherently} as a result of the geometry of the space-time.

Clearly we need a systematic way to characterize the relationship
between  rare events like this and their origin through either
non-Gaussianity or global anisotropy. Analysis of the temperature
patterns using standard descriptors in non-standard scenarios will
produce signals different from what one would see in the presence of
stationary Gaussian noise. To illustrate this issue we study the
pixel distribution function of temperature values. This is an
obvious way to test for non-Gaussianity in a random field of
temperature values, but a coherent fluctuation field also possesses
a one-point distribution that yields some diagnostic information. In
this sense, all the Bianchi models are inherently non-Gaussian but
their non-Gaussianity is simply a manifestation of the presence of
anisotropy.

\begin{figure}
\begin{centering}
  \includegraphics[scale=0.4]{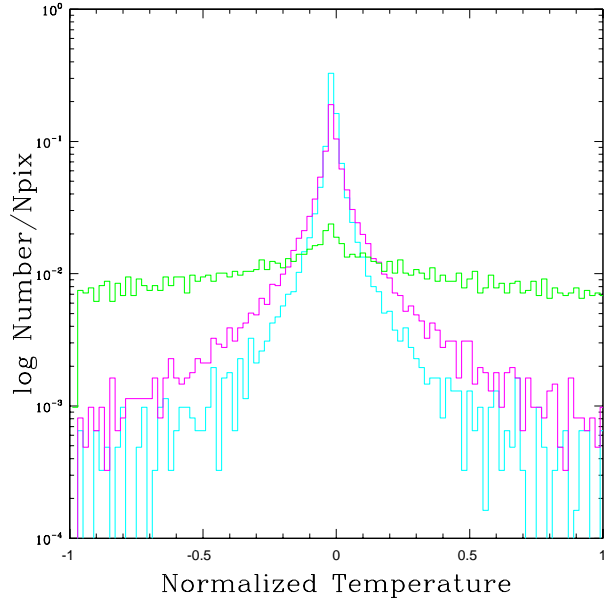}
  \caption{The pixel distribution function of the normalized temperature for the Bianchi CMB maps. The y axis indicates the normalized pixel number which is the whole range of temperature covered by the plot, but normalized so they can be plotted on the same axes. The colour cyan represents type V, magenta for VII$_h$, and green for VII$_0$. The plot shows that the Bianchi V and VII$_h$ types have strongly peaked features at the mean, whereas the type VII$_0$ has a nearly uniform distribution across the temperature range. }
  \label{PDF0}
\end{centering}
\end{figure}

We calculated the pixel distribution function, which is simply a
frequency count of the pixel (or temperature) values, for each of
the Bianchi maps and used it to plot the histogram seen in Figure
\ref{PDF0}. A perfectly homogeneous and isotropic map, such as that
predicted by the concordance model, would have a constant value over
the whole map - remember we are considering maps without
fluctuations - and therefore a histogram of the pixel distribution
histogram for this map would give a delta function at the mean. Our
results show some deviation from this prediction. In Figure
\ref{PDF0} we see the plots for the Bianchi V and VII$_h$ types have
strongly peaked features at the mean, but still with some non zero
variance; the type VII$_0$ model has a nearly uniform distribution
across the whole temperature range.

Although not demonstrated in this diagram, another point to note is
that at early times the histograms of the pixel distribution
functions of the three different Bianchi types are almost identical;
the different values of the temperature pixels are roughly even over
the range. As redshift\footnote{Note, redshift is defined here as
proportional to the inverse geometric mean of three scale factors.}
decreases, the temperature patterns for Bianchi V and VII$_h$
gradually start to focus \citep[see][]{Sung2009}, and their
histograms of the pixel distribution functions become successively
more peaked. For the Bianchi type VII$_0$ the temperature pattern
just twists, reorganizing the pattern on the sky while the histogram
stays roughly the same. These observations help to explain the
features in the histograms. As the temperature patterns become more
focused, more of the rest of the map becomes uniform and so the
histogram is tending towards a delta function.

In summary, what we would hope to discover from the pixel
distribution   histogram is that it gives us some clues about the
homogeneity and isotropy of the maps i.e. the degree of
concentration around the mean might tells us about the homogeneity
of the parameters or the asymmetry of the distribution might give
information about the anisotropy. But as it stands, the information
from the pixel distribution function is not that clear. All we can
say is that the histograms differ from a delta function, so the maps
are not perfectly homogeneous and isotropic, and that the histograms
are clearly non-Gaussian in shape. However the shape of this
one-point pixel distribution does not furnish us with a complete
description of the pattern because it does not take into angular
correlations between the pixels.

\section{Phase correlations of Bianchi CMB Maps}
\label{secPhase}
We now move on from pixel distributions to consider the spherical harmonics of the temperature maps, or more
specifically the phases of the spherical harmonic coefficients.

\subsection{Spherical Harmonics}
\label{secSH} The temperature of the CMB, $\Tp$, is defined on a
sphere  where $\theta \in [0,\pi]$ and $\phi\in [0,2\pi]$ are the
polar and azimuthal angles. Therefore one way of describing the
temperature anisotropies, $\Delta \Tp$, is to extract the
corresponding spherical harmonic coefficients ($\alm$):
\begin{eqnarray}
  \Delta \Tp = \frac{\Tp - \bar T}{\bar T} = \suml \summ |\alm| e^{i\plm} \ylm(\theta,\phi),
\label{eqnALM}
\end{eqnarray}
where $ |\alm|$ and $\plm$ are the amplitudes and  phases of the
spherical harmonic coefficients, and $\ylm$ are the spherical
harmonics which are defined here as:
\begin{eqnarray}
  \ylm = (-1)^{m} \sqrt{\frac{2\l+1}{4 \pi} \frac{(\l-m)!}{(\l+m)!}} P_{\lm} (\cos \theta) e^{im\phi} ,
\label{eqnYLM}
\end{eqnarray}
where $P_{\lm}$ is the associated Legendre Polynomial. Note that
this definition of spherical harmonics includes a phase factor of
$(-1)^{m}$, also known as the Condon-Shortley phase.

In the standard cosmological model, the temperature fluctuation
field is produced by stochastic fluctuations which are Gaussian and
statistically stationary over the celestial sphere. In this case the
phases $\plm$ of each spherical harmonic mode $\alm$ are independent
and uniformly random on the interval $[0,2\pi]$ \citep{Coles2004}.
If instead the temperature pattern on the sky is produced by a
Bianchi geometry then the $\alm$ are no longer stochastically
generated but can be directly calculated from parameters of the
model. Analytical forms for the temperature pattern can be used to
obtain the spherical harmonic phases \citep{McEwen1,McEwen2}, but it
is clumsy to transform these between different coordinate systems
\citep{Coles2004}. In the following we therefore obtain
distributions of $\plm$ from Bianchi maps generated using
the method described by \cite{Sung2010}.

\subsection{Visualising phase correlations}
\label{secVPC} 
\begin{figure}
\begin{centering}
  \includegraphics[width=60mm]{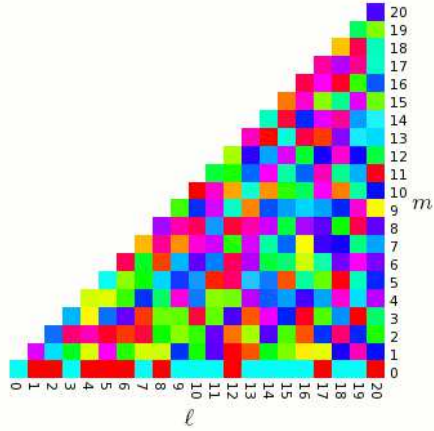}
  \caption{Example of the spherical harmonic phases ($\plm$) we would expect for the concordance model where $\l$, $m$ $\in$ [0, 20]. The distribution of $\plm$ is random, apart from for $m=0$, where the phases can only be 0 or $\pi$ by definition. The colours represent the different values of the $\plm$ (red ($\plm = 0$), green ($\plm =\pi/2$), cyan ($\plm =\pi$), and purple ($\plm=3\pi/2$)). }
  \label{figRandom}
\end{centering}
\end{figure}

To visualize the information held in the phases,
$\plm$, of the spherical harmonic coefficients, $\alm$, we plotted
them over all $\l$ and $m$. Rather than using a 3D plot, colour has
been used to represent the $\plm$ following \cite{cc2000}. The
colours equate to the angle on a colour wheel: red ($\plm = 0$),
green ($\plm =\pi/2$), cyan ($\plm =\pi$), and purple
($\plm=3\pi/2$). To understand these plots, first consider
what we would expect to see in the case of an isotropic and
homogeneous universe as predicted by the concordance model. This
would be a uniform map (as we are not at this point considering
fluctuations) but in spherical harmonics this only has power in one
mode ($\l=m=0$), so there is no phase for the other modes. Better to
consider a map with Gaussian fluctuations as later in the section we
will move on to add noise to the Bianchi maps. Figure
\ref{figRandom} shows the phases ($\plm$) for a homogeneous and
isotropic map with Gaussian fluctuations. The phases are random over
the space i.e. there are no visible patterns in the distribution of
colours in the plot. Note that for all the maps, $\plm$ = 0 or $\pi$
for $m$ = 0 because the $\alm$ coefficients are defined so that
$\alm=a_{l,-m}$. Other than this, the distribution of $\plm$ is
random. Also, note that for all $|m| > \l$, $\plm$ = 0.

\begin{figure}
\begin{centering}
  \includegraphics[width=41mm]{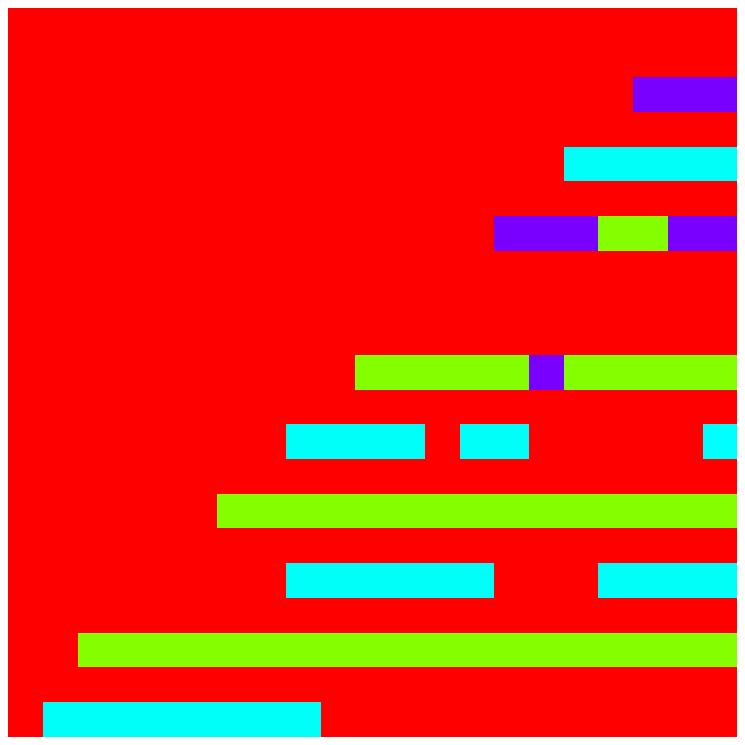}
  \includegraphics[width=41mm]{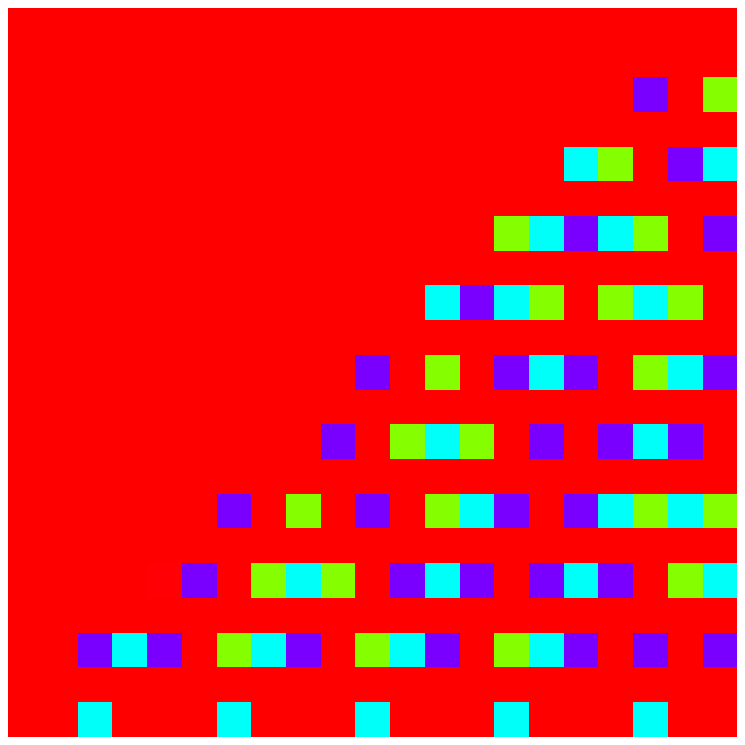}
  \includegraphics[width=41mm]{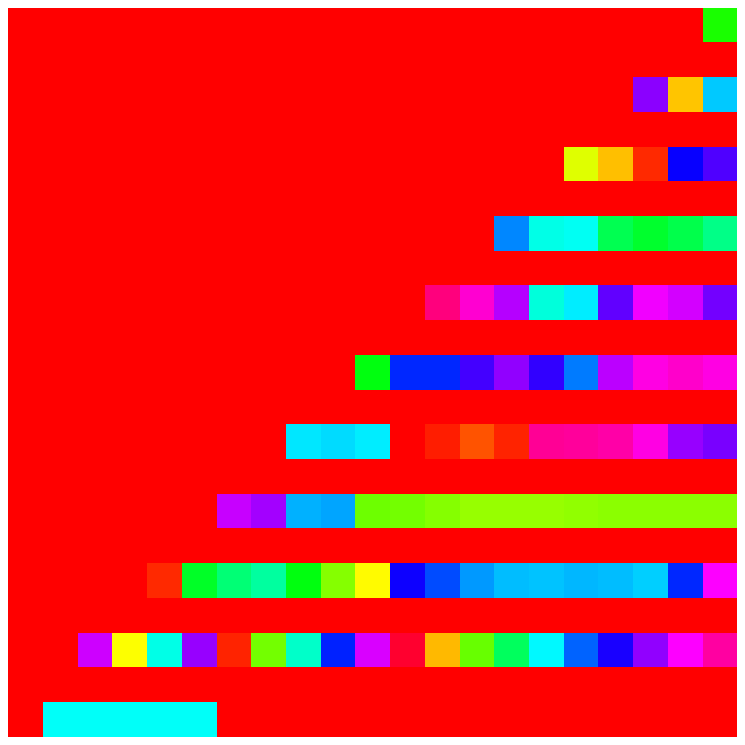}
  \caption{Phases of the spherical harmonic coefficients ($\plm$) for Bianchi types V (left), VII$_0$ (middle) and VII$_h$ (right) where $\l$, $m$ $\in$ [0, 20] and z = 0. Note that $\l$ is plotted against the x axis, increasing from left to right, and $m$ is plotted against the y axis, increasing from bottom to top. The distributions are not random (as in Figure \ref{figRandom}) but exhibit some distinctive features. All the $\plm$ for the VII$_0$ and V types are orthogonal, and there are sequences of colours in the type VII$_h$ (see $m=2$).}
  \label{figPhase}
  \includegraphics[width=41mm]{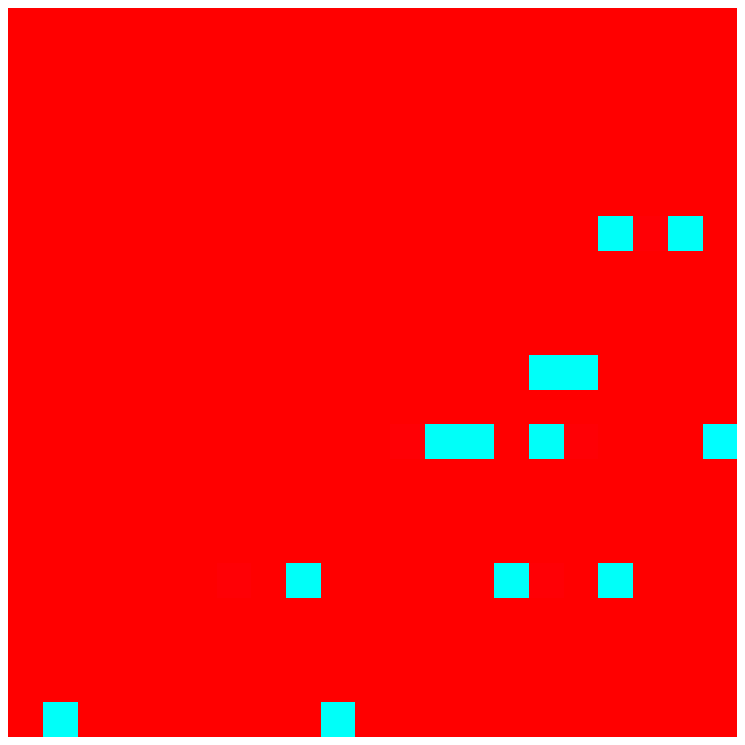}
  \includegraphics[width=41mm]{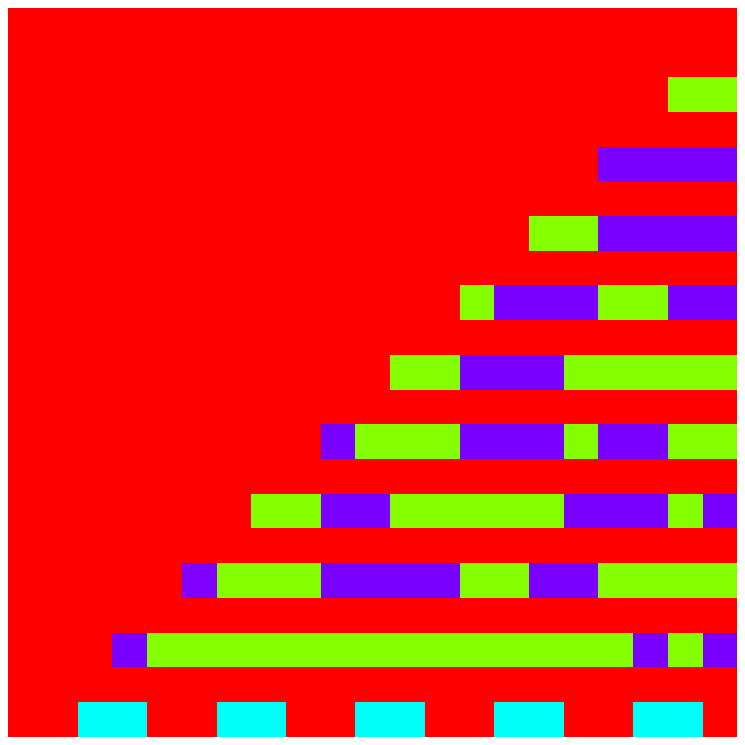}
  \includegraphics[width=41mm]{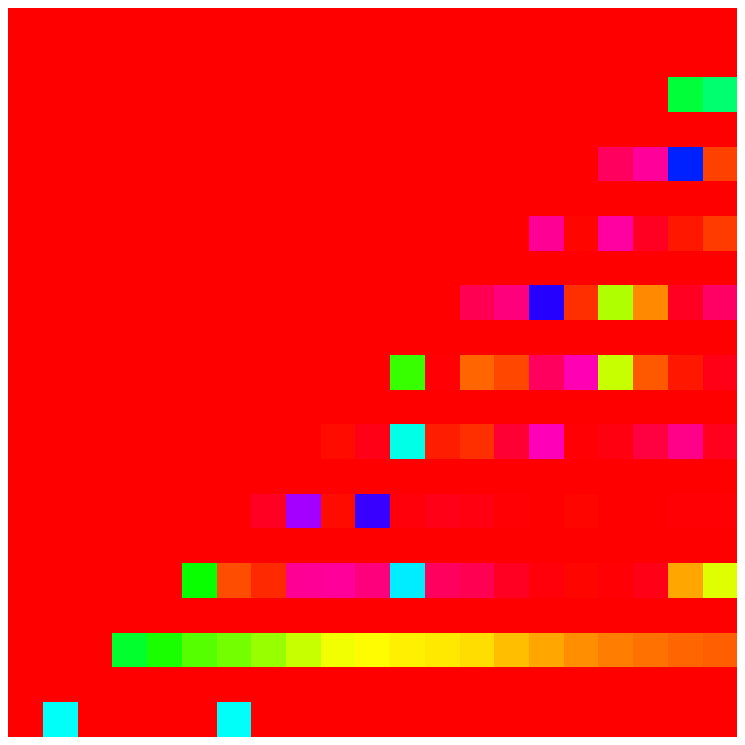}
  \caption{$\Delta\plm$ for Bianchi types V (left), VII$_0$ (middle) and VII$_h$ (right) where $\l$, $m$ $\in$ [0, 20] and z = 0. Note that $\l$ is plotted against the x axis, increasing from left to right, and $m$ is plotted against the y axis, increasing from bottom to top. Like the phases (Figure \ref{figPhase}) the distributions are not random but exhibit some distinctive features. All the $\Delta\plm$ for the V type are either 0 or $\pi$. The $\Delta\plm$ for the VII$_0$ type are again orthogonal but in a more correlated way. Similarly, the sequences of colours in the type VII$_h$ are now even more prominent (see $m=2$).}
  \label{figPhaseDif}
\end{centering}
\end{figure}

We extracted the $\plm$ from each of the Bianchi maps using \healpix
\citep{heal} and plotted them in the same way as we have described;
the results are shown in Figure \ref{figPhase}. The plots show
that the $\plm$ are not random but have patterns, i.e. the harmonic
modes manifest some form of phase correlation. For all the Bianchi
types, $\plm$ = 0 for all odd $m$. For the VII$_0$ and V types, all
the $\plm$ are orthogonal i.e. they are either 0, $\pi/2$,
$\pi$, or $3\pi/2$. Both the VII$_0$ and VII$_h$ types show
sequences of increasing/decreasing phases, which are particularly
prominent for $m=2$.

While some patterns are apparent in these plots, an even better way
to visualize the phase correlations is to look at the phase
differences which are defined here as:
\begin{eqnarray}
  \Delta \plm = \plm -\Phi_{\l,m-1}
\end{eqnarray}
The phase differences are shown in Figure \ref{figPhaseDif}  and the
correlations are much more apparent compared to the plots of $\plm$.
All the $\Delta\plm$ for the V type are lined up, i.e. either 0 or
$\pi$. The $\Delta\plm$ for the VII$_0$ type are again orthogonal,
but whereas in the phases the distribution of 0, $\pi/2$,
$\pi$, and $3\pi/2$ seemed some what random, in the phase
differences we see similar values `clump' together. Similarly, the
sequences of colours in the type VII$_h$ (see $m=2$ for example) are
now even more prominent.

So, strong correlations are observed in the phases and phases
differences  of the simulated Bianchi CMB maps. But we have only
looked at large angular scales where there are only a small number
of independent data points. Even without noise, it is important to
ask the question whether these correlations are likely to be
statistically significant. One way to quantify this is to use a
\textit{Kolmogorov-Smirnov test}. This is a non-parametric
statistical test which measures the maximum distance of a given
distribution from a reference probability distribution. In this case
we want to show deviation from a random set of $\Delta \plm$, i.e. a
uniform distribution, which is predicted by the concordance model.

To calculate the Kolmogorov-Smirnov test statistic a set of  phase
differences $\Delta \plm$ are separated into bins of equal size
between 0 and $2\pi$. The number of $\Delta \plm$ which fall into
each bin are counted and a cumulative distribution derived. If the
distribution is uniform, as in the case of the reference probability
distribution, then the number of $\Delta \plm$ in each of the bins
should increase roughly linearly across the bins. The difference
between both the sample and uniform cumulative distributions is
found for each bin and the biggest difference is the
Kolmogorov-Smirnov statistic \textbfit{D}.

To deduce the significance of \textbfit{D}, a set of ten  thousand
tests have been run to generate sets of random angles of equal size
to the sample sets. \textbfit{D} was found for each of these sets
and this data was used to find the significance of \textbfit{D} for
the sample distributions from the Bianchi maps.

The Kolmogorov-Smirnov statistic \textbfit{D}, and the  derived
probability of that statistic P(\textbfit{D}), for all the Bianchi
maps are detailed in Table \ref{tableKS}.
\begin{table}
\begin{centering}
  \begin{tabular}{|l|l|l|l|l|l|}
    \hline
    Map & z &  \textbfit{D} & P(\textbfit{D}) \% \\ \hline
    VII$_h$ & 500  & 0.11  &     94.5        \\
    VII$_h$ & 200  & 0.09  &     77.1        \\
    VII$_h$ &  60  & 0.14  &     99.2        \\
    VII$_h$ &  10  & 0.24  &  $>$99.9        \\
    VII$_h$ &   3  & 0.27  &  $>$99.9        \\
    VII$_h$ &   1  & 0.38  &  $>$99.9        \\
    VII$_h$ &   0  & 0.27  &  $>$99.9        \\
    VII$_0$ &   0  & 0.28  &  $>$99.9        \\
    V       &   0  & 0.73  &  $>$99.9        \\
    \hline
  \end{tabular}
  \caption{Results from the Kolmogorov-Smirnov test comparing the distribution of phase differences in the Bianchi CMB maps with a random distribution of phases as predicted by the concordance model. \textbfit{D} is the Kolmogorov-Smirnov statistic found by comparing the phase differences $(\Delta\Phi$). $P(\textbfit{D})$ is the Monte Carlo estimate of the probability of getting the value of \textbfit{D}, or less, found for the Bianchi models, from a random selection of phase differences. These are computed by forming an empirical distribution of \textbfit{D} from sets of random simulations and counting what fraction of the ensemble gives the results obtained for the Bianchi maps. For example, in the case of the $P(\textbfit{D})$ for the VII$_h$ map (z = 500) we find that, out of 10000 simulations, 9450 have a value of \textbfit{D} {\em less than} 0.11. Given the probable sampling accuracy of around one percent, we have rounded the results.}
  \label{tableKS}
\end{centering}
\end{table}
This table shows that there is indeed a  significant deviation from
a uniform distribution for the phase differences for all Bianchi
types. Of the 10000 random sets of data, none showed a value for
\textbfit{D} as high as seen for the Bianchi cases.

The Bianchi VII$_h$ type was also considered at different  redshifts
to see how the correlations changed with time. Table \ref{tableKS}
shows that in general value of \textbfit{D} gets more significant
over time i.e. the correlations in the phase differences of the
Bianchi maps become stronger over time.

\subsection{Rotating maps and adding noise}
\label{secRN} In Section \ref{secVPC} we applied the
Kolmogorov-Smirnov test to a ``clean'' map that is perfectly aligned
with the vertical axis. This section addresses how noise and
rotation affect the identification of correlations in the phases of
the spherical harmonics of CMB maps from Bianchi models.

First we consider rotation. Phases of spherical harmonic coefficients are not rotation-invariant. Rotating the coordinate system used to represent a CMB map in $\phi$ (which is equivalent to rotation around the z axis) would increment each of the spherical harmonic phases by $\phi$, so the phase differences would remain the same. Therefore rotation in $\phi$ would have no effect on the value of the Kolmogorov-Smirnov statistic \textbfit{D}. Rotation in $\theta$ is more complicated to express so we used an empirical approach to quantity the effect on \textbfit{D}. The Bianchi CMB maps were rotated by a small angle, $\theta = \pi/8$, and then the spherical harmonic coefficients were derived and used to calculate \textbfit{D}. The results in Table \ref{tableKSRot} show that the values of \textbfit{D} for each of the maps are even higher than in maps that hadn't been rotated, indicating the presence of even stronger correlations. This suggests that, at least for small rotations off the axis, the correlations are just as significant, if not more so.
\begin{table}
\begin{centering}
  \begin{tabular}{|l|l|l|l|l|l|}
    \hline
    Map & z &  \textbfit{D} & P(\textbfit{D}) \% \\ \hline
    VII$_h$ &   0  & 0.46  &    $>$99.9 \\
    VII$_0$ &   0  & 0.38  &    $>$99.9 \\
    V       &   0  & 0.77  &    $>$99.9 \\
    \hline
  \end{tabular}
  \caption{Results from the Kolmogorov-Smirnov test comparing the distribution of phase differences in the Bianchi CMB maps, rotated by $\theta = \pi/8$, with a random distribution of phase differences as predicted by the concordance model. \textbfit{D} is the Kolmogorov-Smirnov statistic found when considering the phase differences. $P(\textbfit{D})$ is the Monte Carlo estimate of the probability of getting the value of \textbfit{D}, or less, found for the Bianchi models, from a random selection of phase differences. These are computed by forming an empirical distribution of \textbfit{D} from sets of random simulations and counting what fraction of the ensemble gives the results obtained for the Bianchi maps. For example, in the case of the $P(\textbfit{D})$ for the VII$_h$ map we find that, out of 10000 simulations, over 9999 have a value of \textbfit{D} {\em less than} 0.46.}
  \label{tableKSRot}
\end{centering}
\end{table}

As an aside, the colour plots of the phase differences for  Bianchi
maps rotated by a number of different $\theta$ in the range 0 to
2$\pi$ were generated. These plots have been condensed together into
movies\footnote{The movies can be found at
http://www.astro.cardiff.ac.uk/research/theoreticalcosmology/?page=research}
which show that the correlations in the VII$_0$ and V maps are
visible across all $\theta$ and for the VII$_h$ map are visible within
about $\pi/3$ of the preferred axis. .
\begin{figure}
\begin{centering}
  \includegraphics[width=41mm]{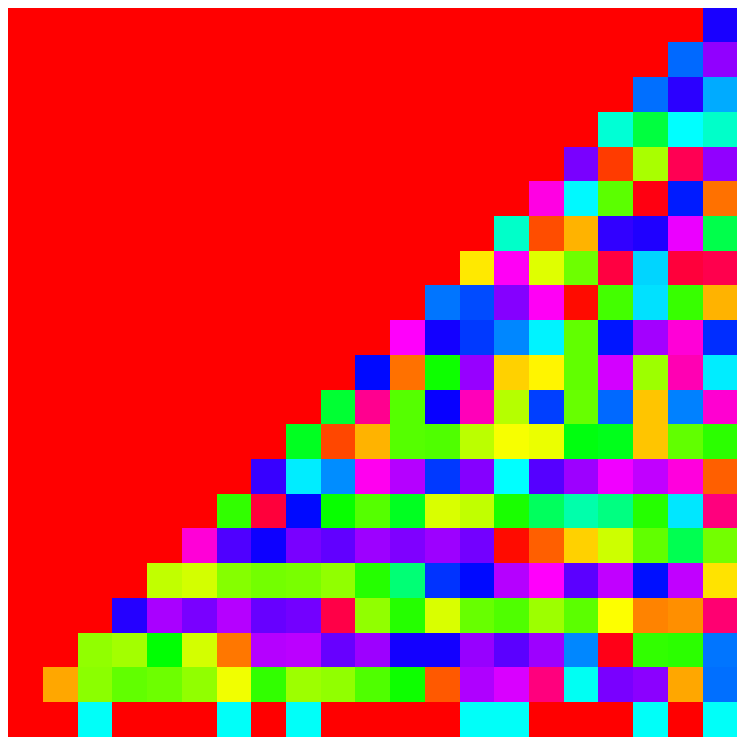}
  \includegraphics[width=41mm]{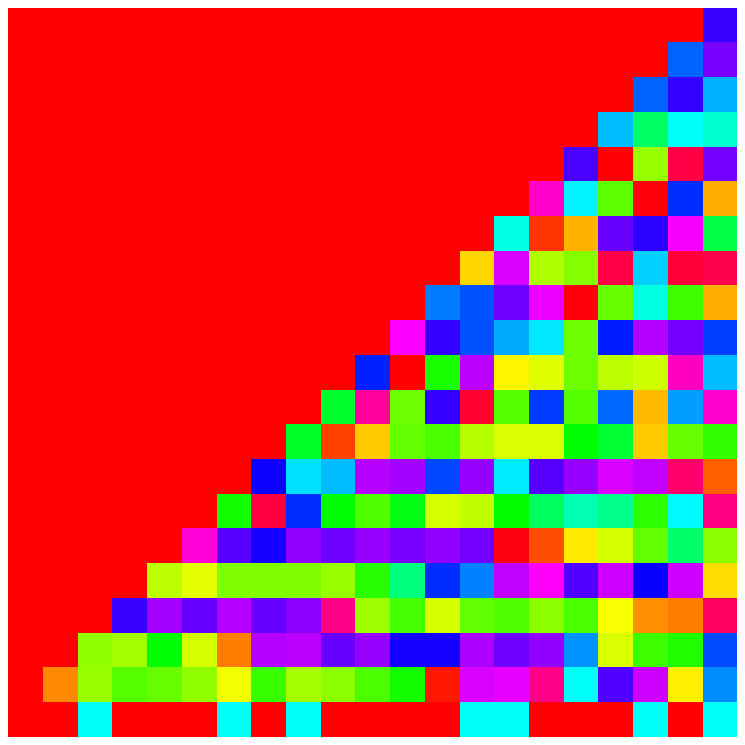}
  \includegraphics[width=41mm]{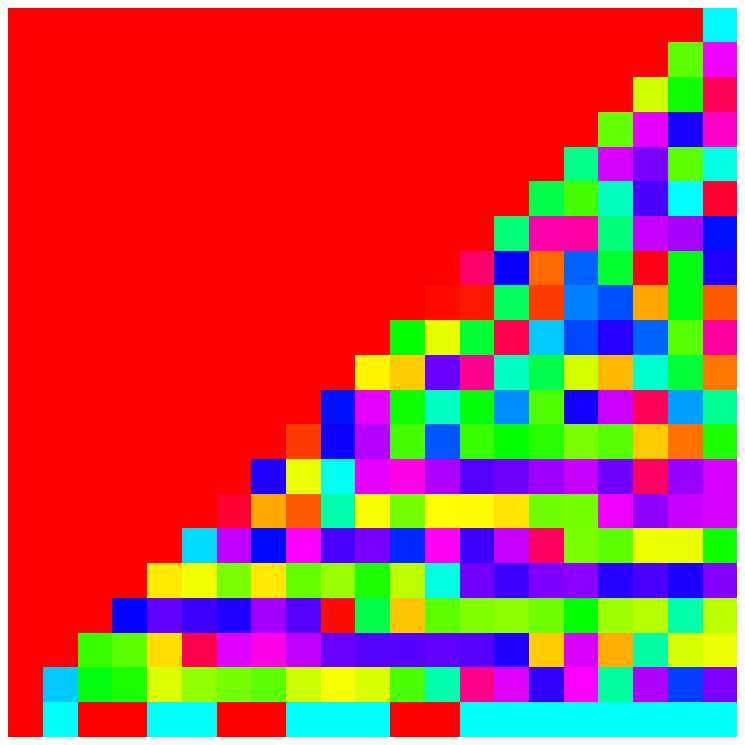}  
  \caption{$\Delta\plm$ for $\l$, $m \in$ [0, 20] (Bianchi type V map at z = 0 with white (left), WMAP (middle), and $\Lambda$CDM fluctuations (right) noise maps ,  $\theta=\pi/8$). Note that $\l$ is plotted against the x axis, increasing from left to right, and $m$ is plotted against the y axis, increasing from bottom to top. Correlations can still be observed as lines of similar colours.}
  \label{figRotNoisePhaseDif}
\end{centering}
\end{figure}
\begin{table}
\begin{centering}
  \begin{tabular}{|c|c|c|c|c|c|c|c|}
    \hline
    Map & z & \textbfit{D$_{white}$} & P(\textbfit{D})\% & \textbfit{D$_{wmap}$} & P(\textbfit{D})\% & \textbfit{D$_{\Lambda CDM}$} & P(\textbfit{D})\% \\ \hline
    VII$_h$ &  0     & 0.15  &  99.2    & 0.16  &  99.8   & 0.17  &  99.8 \\
    VII$_0$ &  0     & 0.09  &  77.1    & 0.08  &  66.1   & 0.07  &  52.0    \\
    V       &  0     & 0.19  &  $>$99.9 & 0.18  &  99.9   & 0.14  &  98.2  \\
    \hline
  \end{tabular}
  \caption{Results from the Kolmogorov-Smirnov test comparing the distribution of phase differences in the Bianchi CMB maps rotated by $\theta = \pi/8$ with white, WMAP, and $\Lambda$CDM noise maps (z = 0) . \textbfit{D} is the Kolmogorov-Smirnov statistic found when considering the phase differences. $P(\textbfit{D})$ is the Monte Carlo estimate of the probability of getting the value of \textbfit{D}, or less, found for the Bianchi models, from a random selection of phase differences. These are computed by forming an empirical distribution of \textbfit{D} from sets of random simulations and counting what fraction of the ensemble gives the results obtained for the Bianchi maps. For example, in the case of the $P(\textbfit{D})$ for the VII$_h$ map with white noise we find that, out of 10000 simulations, 9920 have a value of \textbfit{D} {\em less than} 0.15.}
  \label{tableKSRotNoise}
\end{centering}
\end{table}

Now to investigate the effect of noise, we considered three different types of noise. Firstly we tried the simplest form by just adding white noise to the Bianchi map. A map of random Gaussian noise (white noise) was generated. Using Healpix the spherical mode resolution was reduced to $\ell$ $\le$ 20. Then the ``noise'' map was modified to have zero mean and variance half that of the Bianchi map. The second ``noise'' map was derived from a product available on the WMAP L\textsc{ambda}\footnote{http://lambda.gsfc.nasa.gov/} website which provides the effective number of observations per pixel. A map of random Gaussian noise was again generated. The variance was modified per pixel so that it was inversely proportional to the square of the number of observations in that pixel. Using Healpix the spherical mode resolution was reduced to $\ell$ $\le$ 20. Then the noise map was modified to have zero mean and variance half that of the Bianchi map.  The final ``noise'' map used a simulation of $\Lambda$CDM fluctuations of the CMB (as performed by \cite{Eriksen2005}). Again the noise map was modified to reduce the spherical mode resolution to $\ell$ $\le$ 20 and have variance half that of the Bianchi map. 

Each of these ``noise'' maps was added to each of the rotated Bianchi maps. We see from the example in Figure \ref{figRotNoisePhaseDif} that the spherical harmonic coefficients derived still have visible correlations in the phases for the Bianchi V map. The results of the Kolmogorov-Smirnov
test (see Table \ref{tableKSRotNoise}) show that the correlations are still detectable and significant for the Bianchi V and VII$_h$ maps but not so well for the VII$_0$ maps. So the method is better for detecting focused features that twisted features.

We see that the effect of adding fluctuations here is not dissimilar to adding just Gaussian noise. The
concordance model predicts fluctuations which are stationary and
Gaussian, as discussed in the Introduction (Section
\ref{secIntro}). Although these fluctuations are correlated on the sky, they have random phases so are incoherent with respect to what our statistic measures.

The ``noise'', or fluctuation, maps are added to the Bianchi maps so that the ratio of the variances is of order unity. However, any ratio is possible; this specific choice is just for illustrative purposes to demonstrate the proposed methods. Nevertheless, if a random-phase (Gaussian) signal is superimposed on the Bianchi template, the phase coherence of the resulting map will still be degraded. If the Gaussian component is too large, the overall map will be indistinguishable from one with purely random phases.

In the examples we have shown, the Gaussian ``noise'' or fluctuation maps are added to the Bianchi maps in such a way that the ratio of the overall variance is of order unity. Our method still functions well with this level of ``contamination'', but if the noise variance is much higher than that of the Bianchi maps the method begins to struggle.

So summarizing, the phases of the spherical harmonic coefficients are a very effective way of identifying focusing 
features in CMB maps, as long as the noise is not excessive, and can be used to give quantifiable
significances. However, like the pixel distributions, the variation
from the expectation of the concordance model only gives us an
indication of non-Gaussianity. It is not clear in what form the
non-Gaussianity occurs, such as an anisotropy or inhomogeneity.
Hence the next section looks at multipole vectors which are built
from spherical harmonic coefficients but can be used to give results
in pixel (as opposed to harmonic) space, which is more meaningful
from the point of view of diagnosing the presence of a preferred
direction.

\subsection{Application to WMAP 5 Year Data}
\label{secWMAP} For pedagogical interest, the methods described in
Section \ref{secVPC} are applied  here to the WMAP 5 year Internal
Linear Combination (ILC) map. The results of the Kolmogorov-Smirnov
test on the ILC map in the galactic coordinate system show very low
significance correlations in $\Delta\Phi$ (see Table
\ref{tableKSRotNoise2}). 

\begin{figure}
\begin{centering}
  \includegraphics[width=41mm]{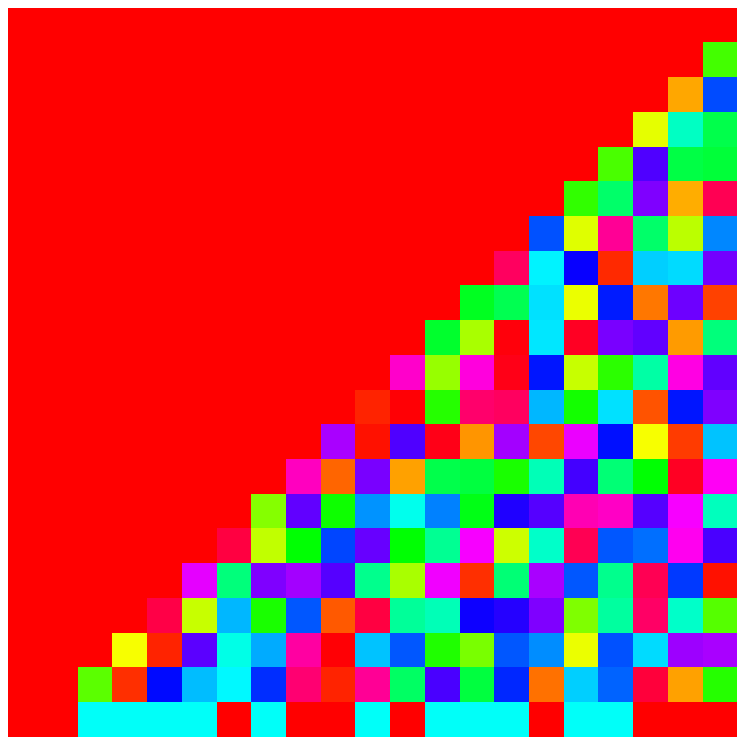}
  \includegraphics[width=41mm]{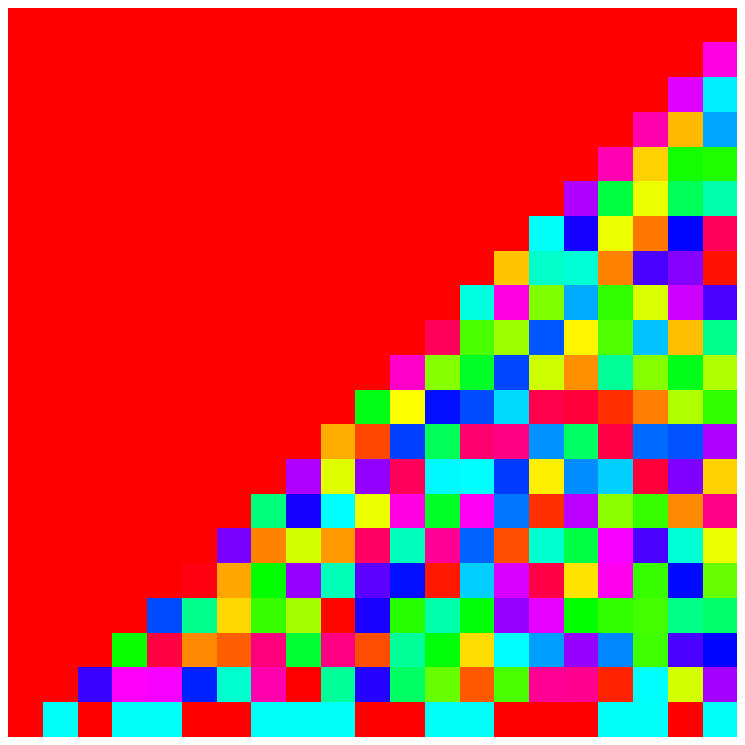}
  \caption{$\Delta\Phi$ for $\l$, $m$ $\in$ [0, 20] for the ILC map with the Axis of Evil aligned with the preferred axis (left) and for the same map but with the phases replaced with random phases (right). Note that $\l$ is plotted against the x axis, increasing from left to right, and $m$ is plotted against the y axis, increasing from bottom to top. No correlations are visible in either plot.}
  \label{figPhaseDifWMAP}
\end{centering}
\end{figure}

However Section \ref{secRN} showed that to
see the phase correlations, the Bianchi maps needed to be rotated
close to the preferred axis. There have been studies that have found
a preferred direction in the WMAP data, highlighted by the alignment
of at least the quadrupole ($\l$ = 2) and octopole ($\l$ = 3). This
preferred axis is known as the Axis of Evil. Therefore the methods
from Section \ref{secVPC} are applied to the ILC map rotated so that
the vertical axis aligns with the Axis of Evil. These $\Delta\Phi$
plotted in Figure \ref{figPhaseDifWMAP} do not show any visual
correlations. For comparison the figure also includes a plot of the
same ILC data but with the phases replaced with random angles (i.e.
so as to not affect the magnitude of the amplitudes of the
$a_{l,m}$). Whilst the Kolmogorov-Smirnov test (Table
\ref{tableKSRotNoise2}) finds higher significance results than when
the map was in Galactic coordinates, the results are still at a low
significance. 

The results show there is no significant detection, so if we do live in an anisotropic universe then it must be obscured with considerable ``noise'' (fluctuations). However the fact that the significance of the results do increase when the map is aligned with the Axis of Evil is intriguing.

\begin{table}
\begin{centering}
  \begin{tabular}{|l|l|l|l|l|l|}
    \hline
    Map     & Axis     & \textbfit{D} & P(\textbfit{D}) \% \\ \hline
    ILC     & Galactic    & 0.06                 &  66.20   \\
    ILC     & Evil        & 0.07                 &  86.05   \\
    \hline
  \end{tabular}
  \caption{Results from the Kolmogorov-Smirnov test comparing the distribution of phase differences in the WMAP ILC map, rotated to align with either the galactic axis or axis of evil, with a random distribution of phase differences as predicted by the concordance model. \textbfit{D} is the Kolmogorov-Smirnov statistic found when considering the phase differences. $P(\textbfit{D})$ is the Monte Carlo estimate of the probability of getting the value of \textbfit{D}, or less, found for the Bianchi models, from a random selection of phase differences. These are computed by forming an empirical distribution of \textbfit{D} from sets of random simulations and counting what fraction of the ensemble gives the results obtained for the Bianchi maps. For example, in the case of the P(\textbfit{D}) for the ILC map in the galactic plane we find that, out of 10000 simulations, 6620 have a value of \textbfit{D} {\em less than} 0.06.}
  \label{tableKSRotNoise2}
\end{centering}
\end{table}

\section{Multipole Vectors from Bianchi Universe}
\label{secMultipole} As shown in the previous section, the
properties of  spherical harmonic coefficients provide us with a
generally effective way of identifying anisotropy through
correlations in CMB maps. However the geometric interpretation of
the mode correlations seen in harmonic space is by no means easy to
interpret geometrically. In an effort to use the spherical harmonics
to provide more meaningful explanation of non-Gaussianities found,
we now consider an alternative approach, based on multipole vectors
These can be constructed from spherical harmonics, using the $\alm$
coefficients derived from CMB maps, but they give results in real
(i.e. pixel) space. For a summary of the main terminology for the
multipole vectors, using the polynomial interpretation approach
which was introduced by \cite{Katz2004}, please see Appendix
\ref{appMV}.

\subsection{Results for Bianchi maps}
\begin{figure}
\begin{centering}
 \includegraphics[scale=0.15,angle=90]{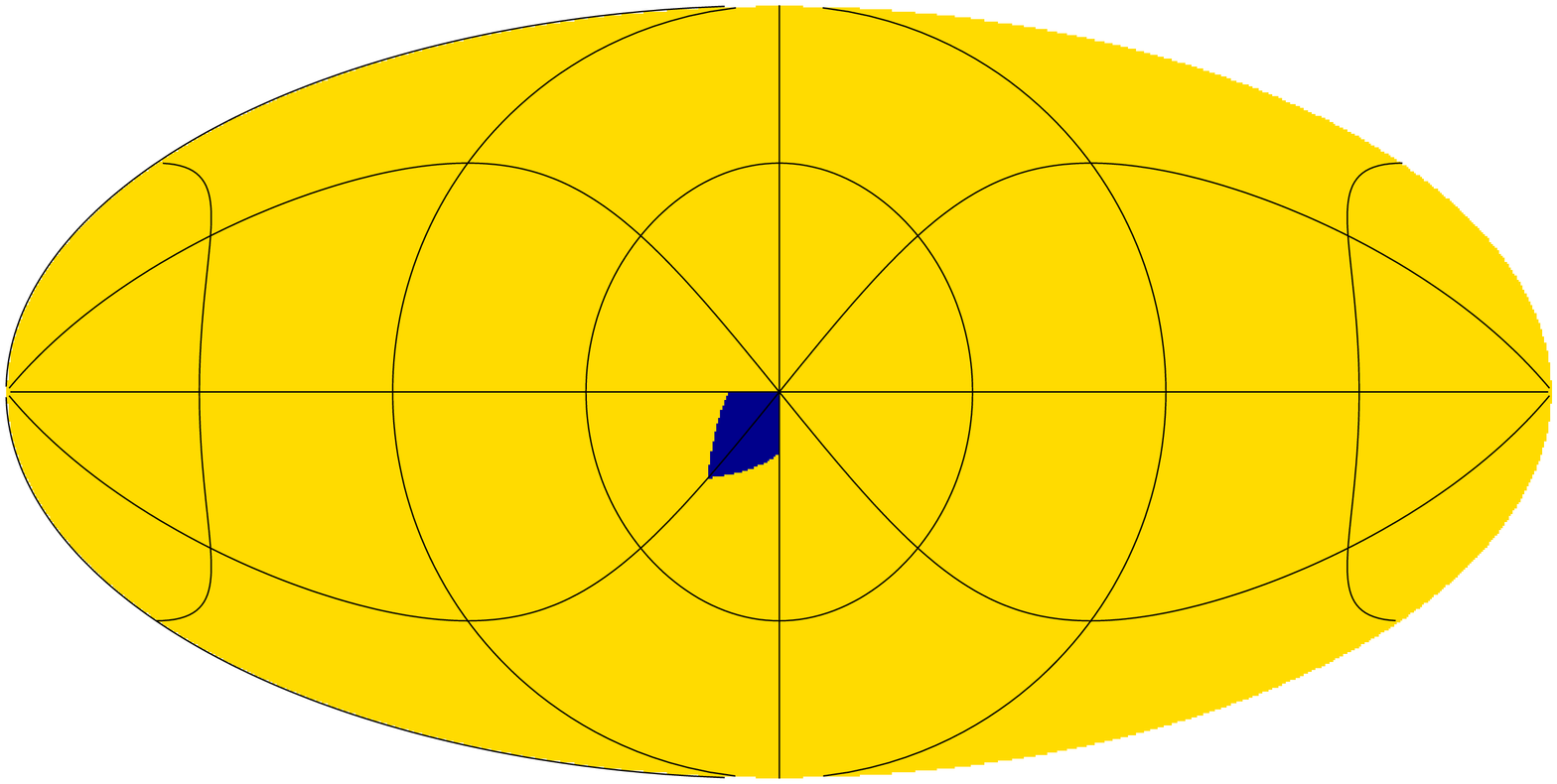}
  \includegraphics[scale=0.15,angle=90]{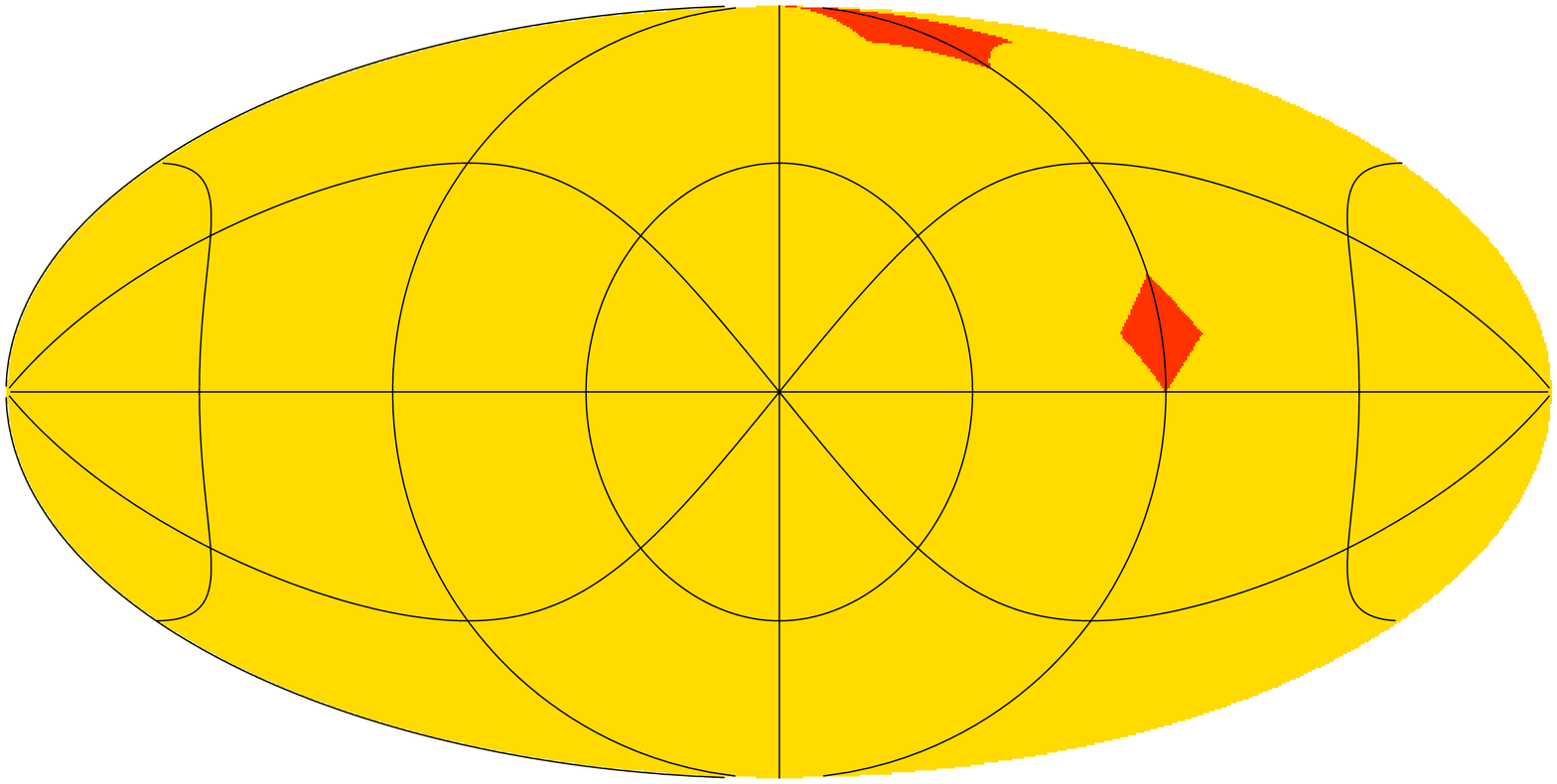}
 \includegraphics[scale=0.15,angle=90]{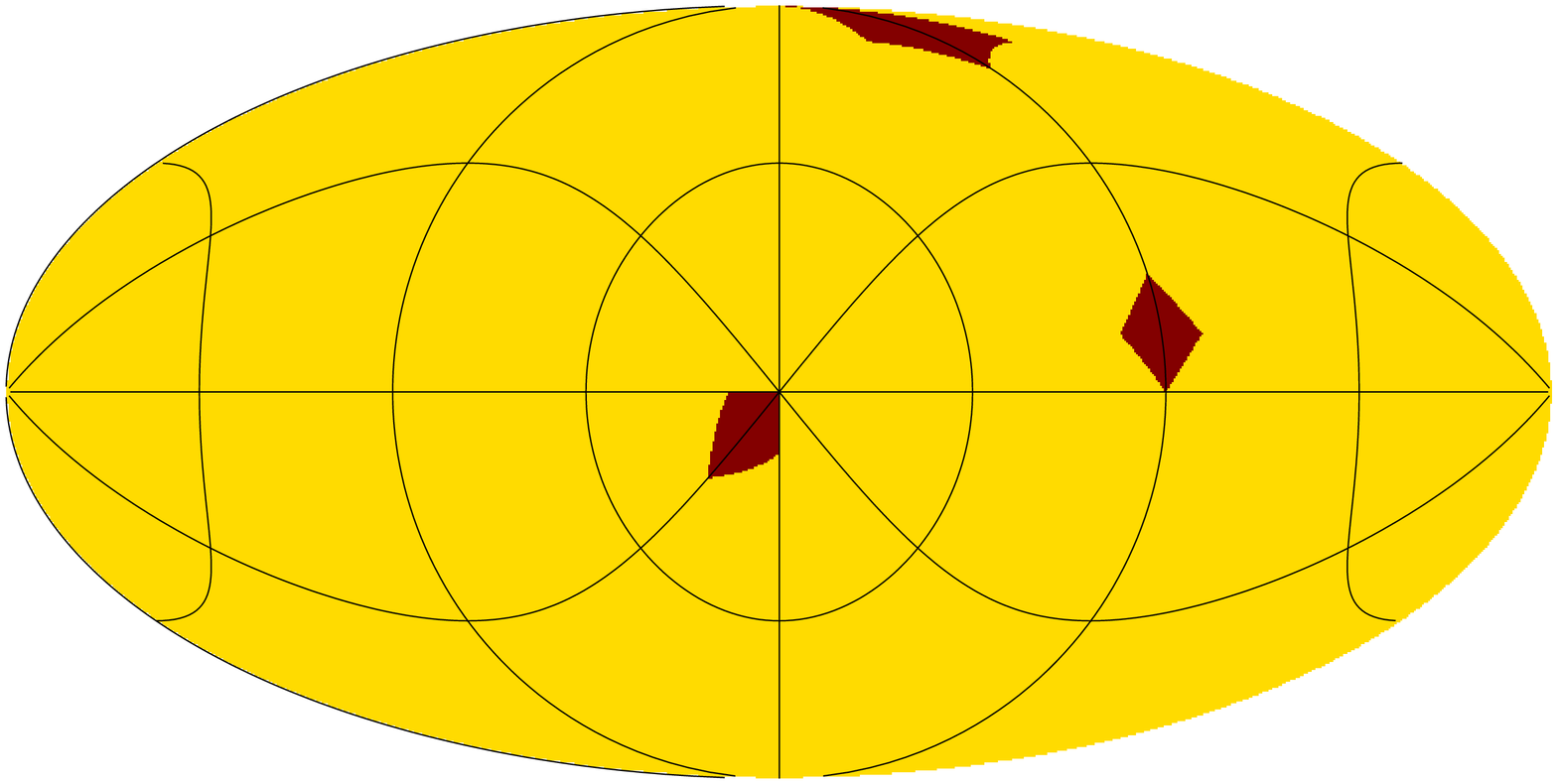}
 \caption{The multipole vectors from Bianchi V map: dipole (left), quadrupole (middle) and octopole (right).}
  \label{MvsVz0}
\end{centering}
\end{figure}
\begin{figure}
\begin{centering}
 \includegraphics[scale=0.15,angle=90]{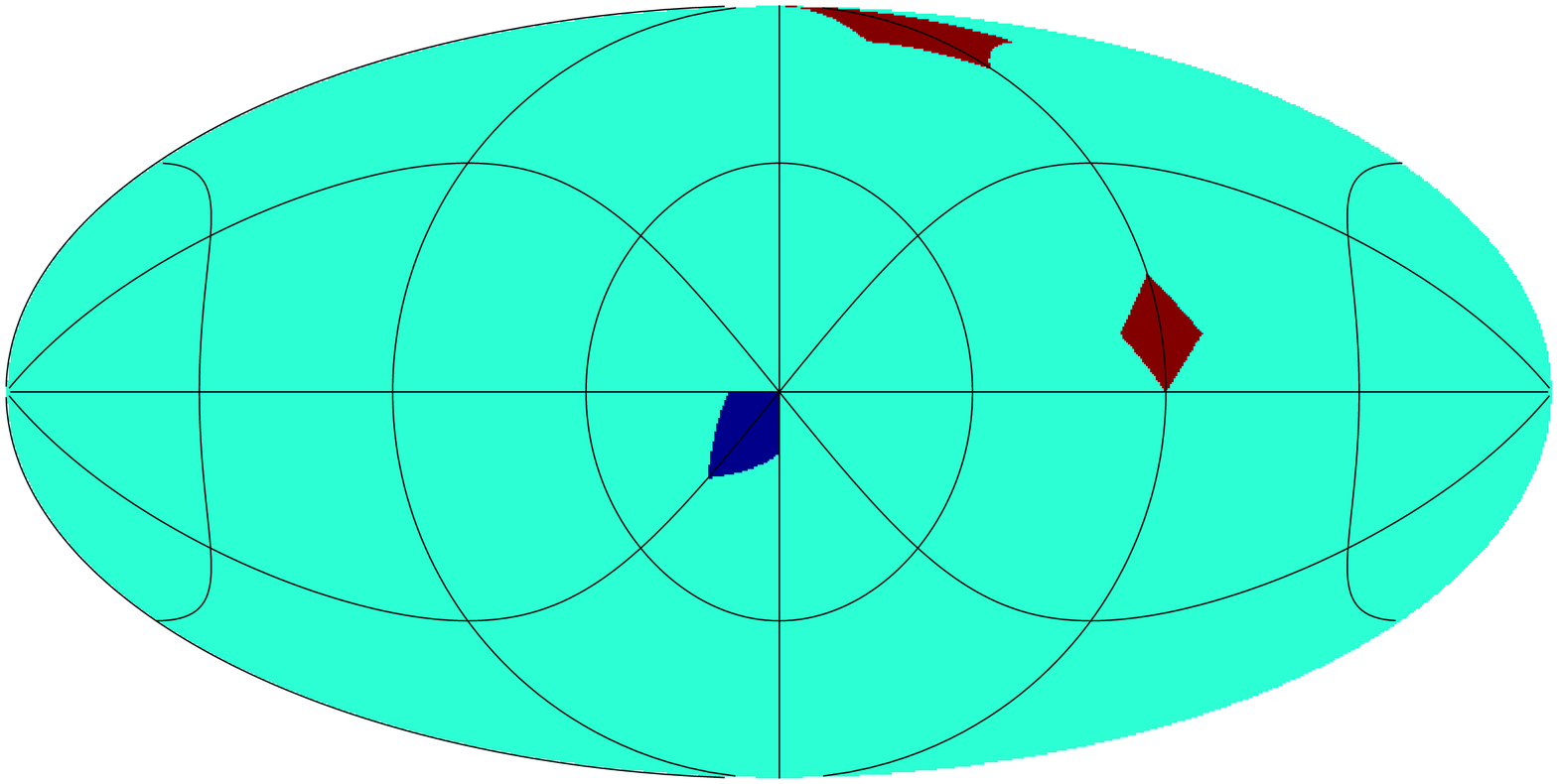}
 \includegraphics[scale=0.15,angle=90]{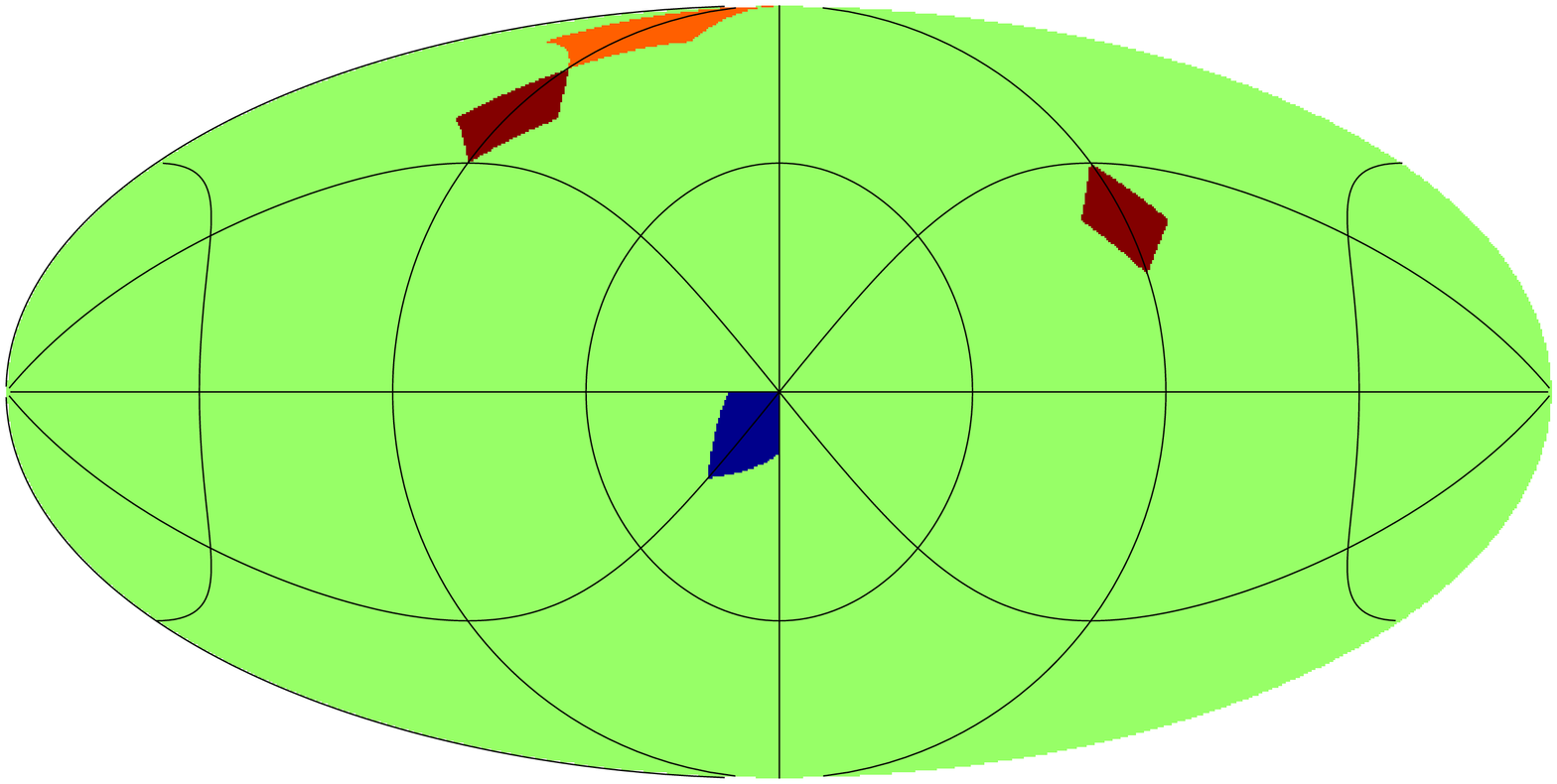}
 \includegraphics[scale=0.15,angle=90]{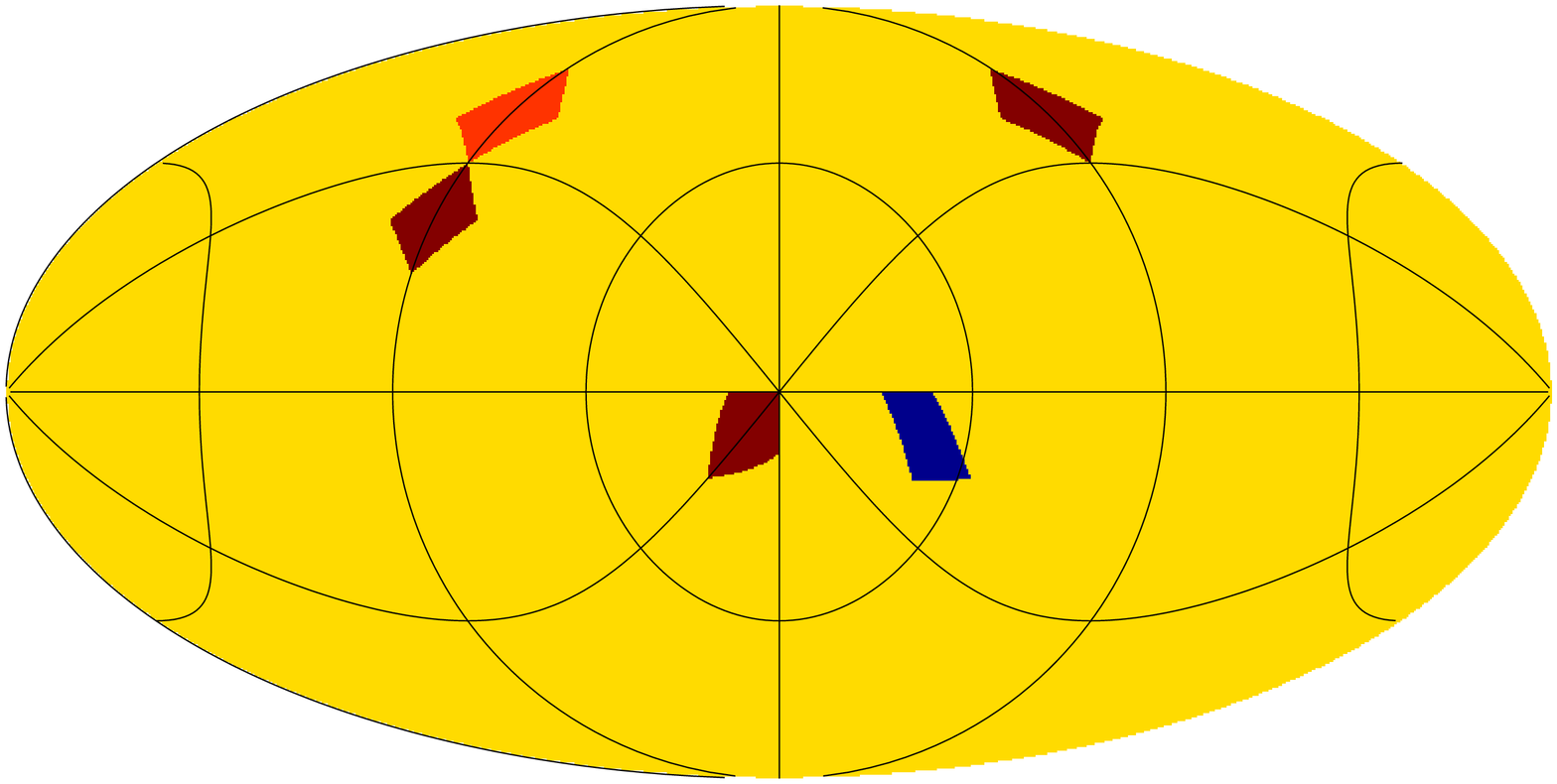}\\
 \includegraphics[scale=0.15,angle=90]{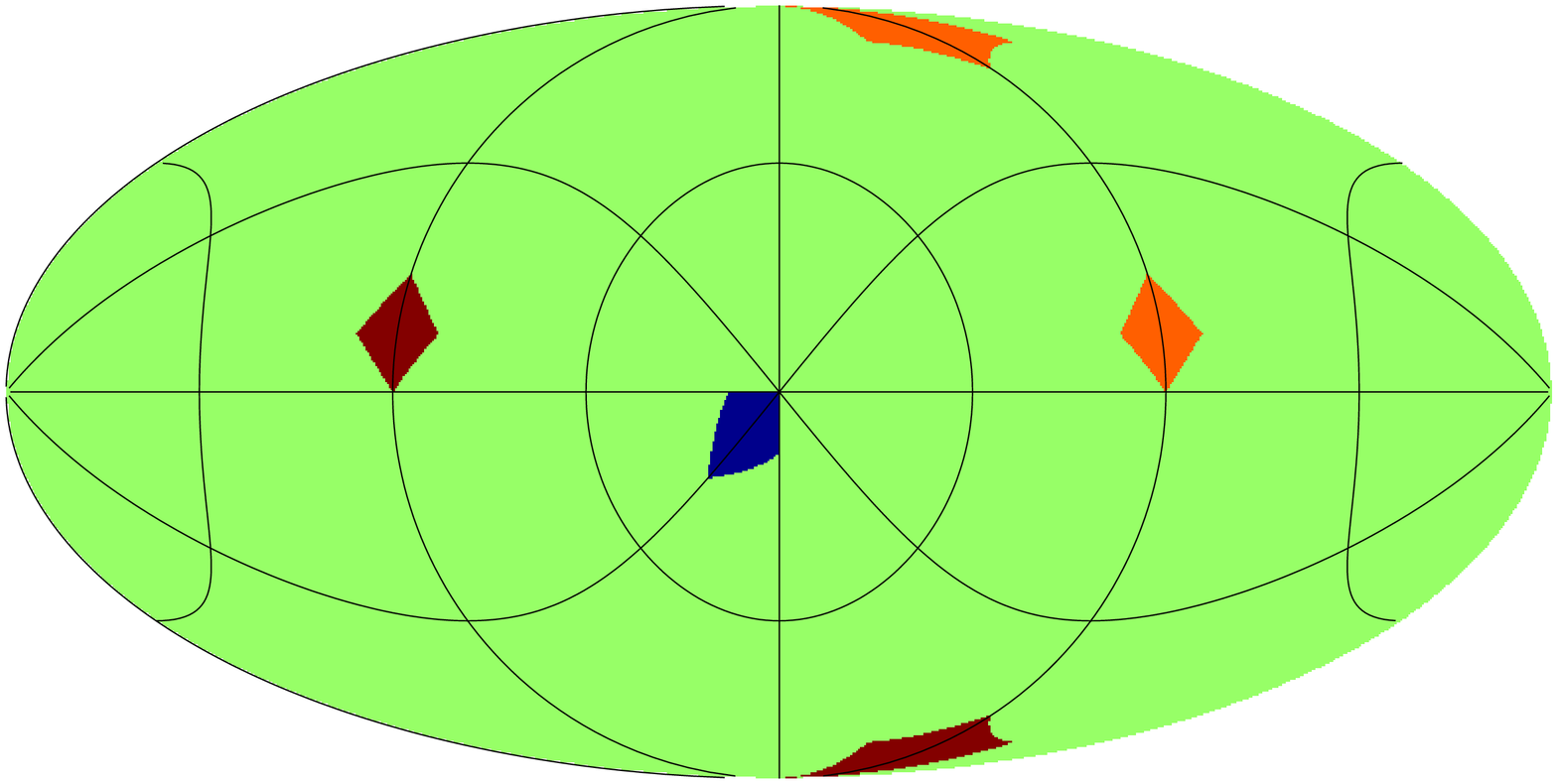}
 \includegraphics[scale=0.15,angle=90]{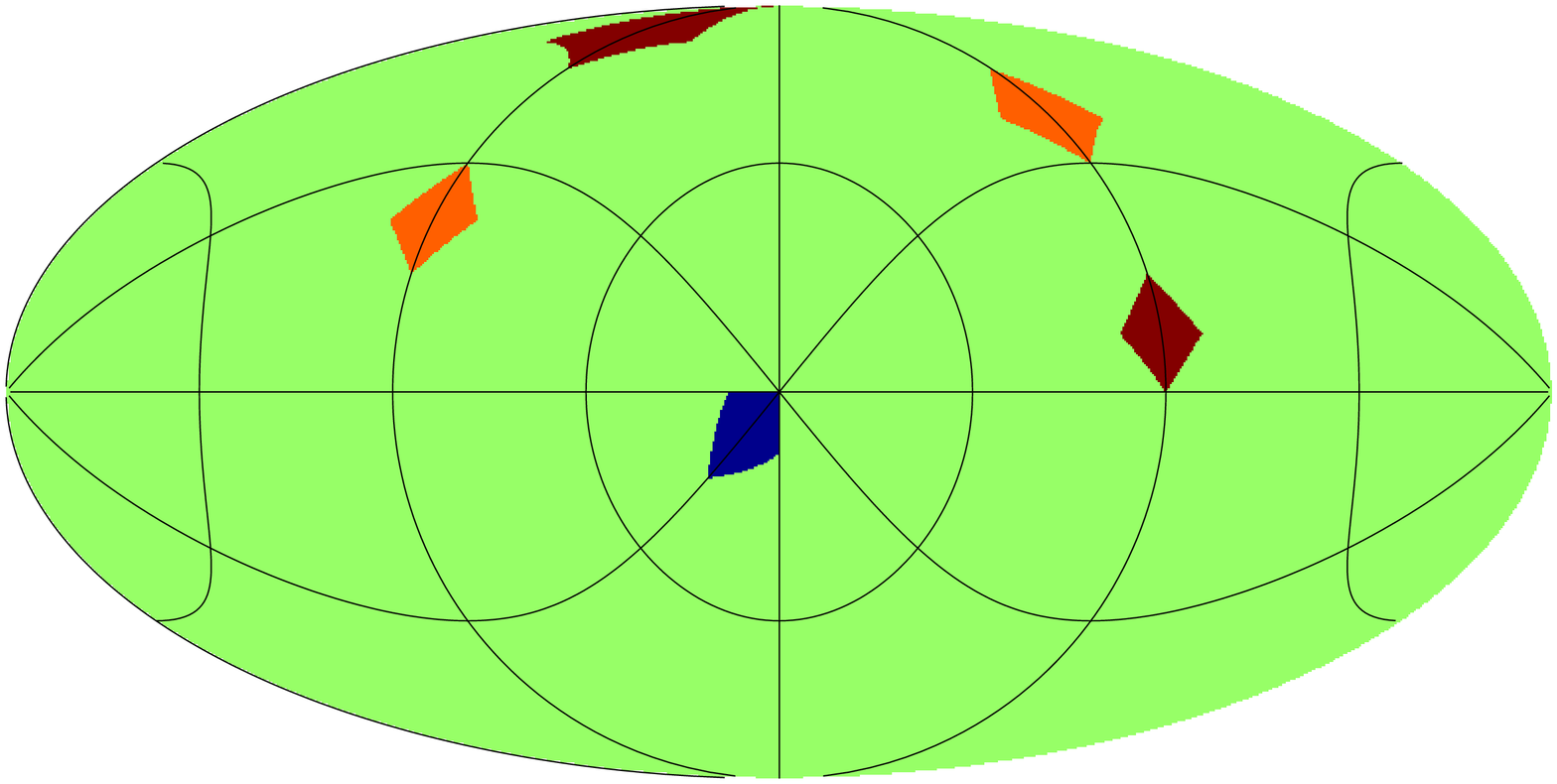}
 \includegraphics[scale=0.15,angle=90]{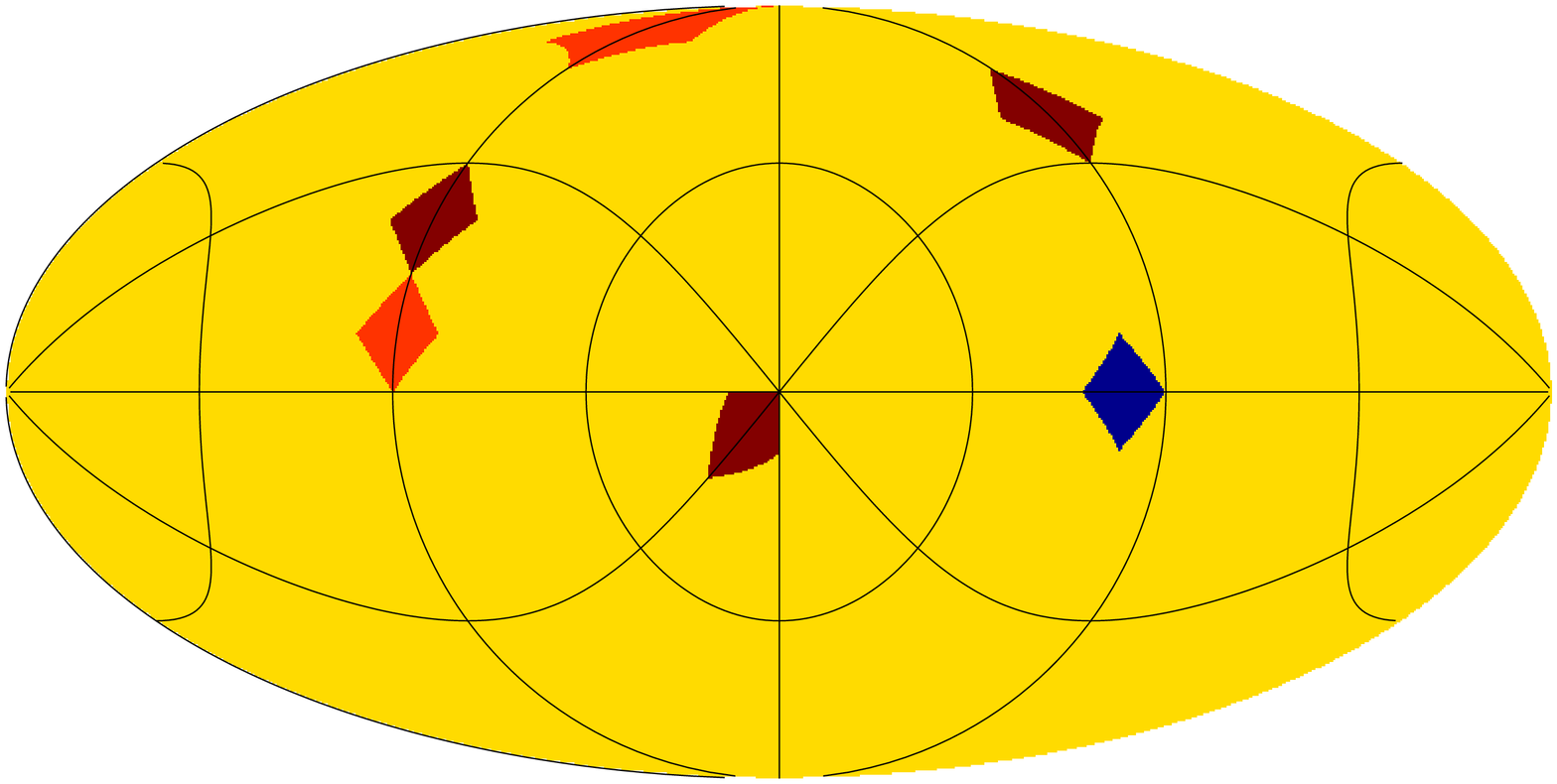}\\
 \includegraphics[scale=0.15,angle=90]{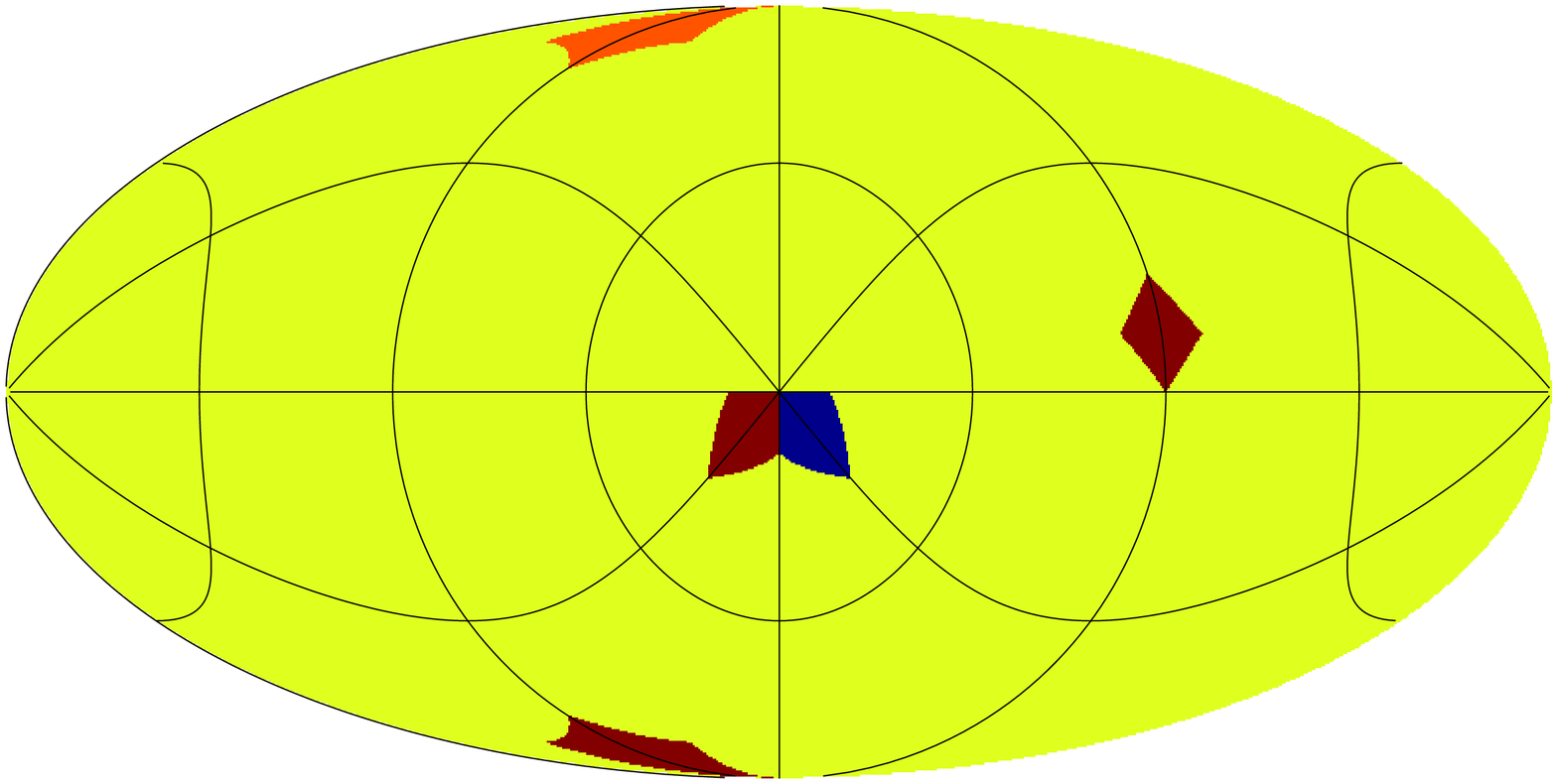}
 \includegraphics[scale=0.15,angle=90]{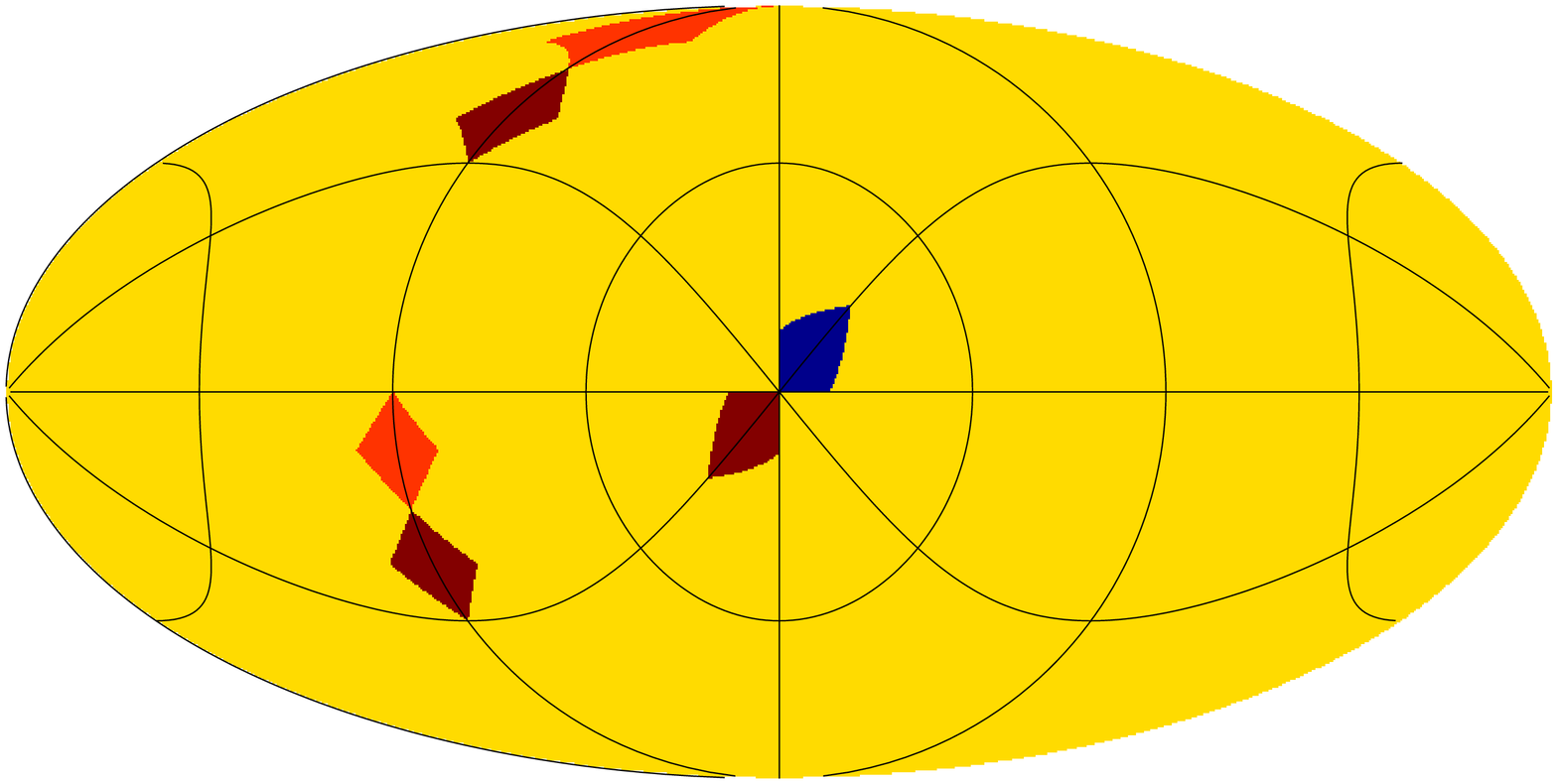}
 \includegraphics[scale=0.15,angle=90]{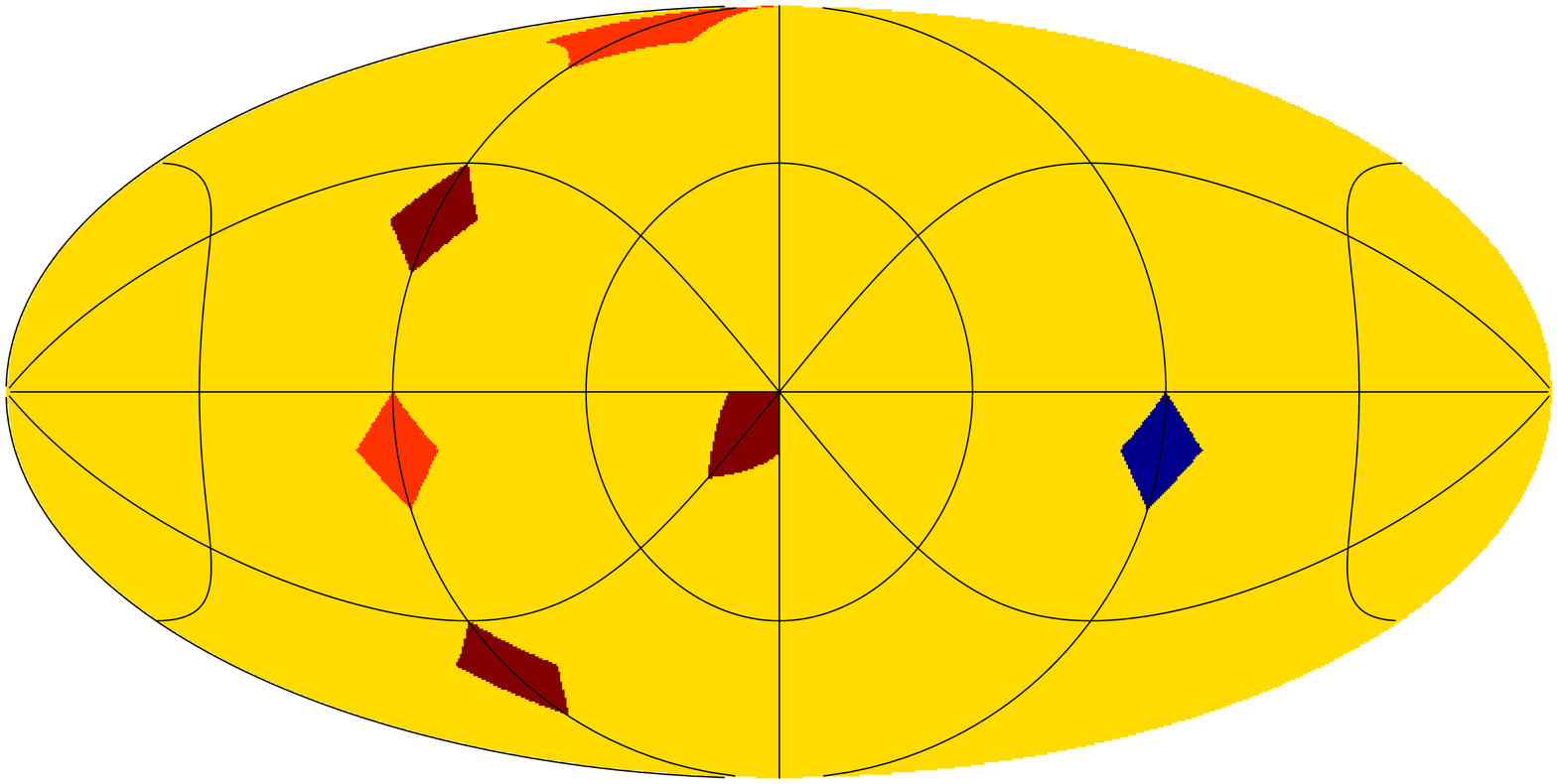}\\
  \caption{The multipole vectors from the Bianchi V (left), VII$_h$ (middle) and VII$_0$ (right) maps from early stage (bottom panel) to late time (top panel). The vectors are represented by dots: dark blue for the dipole, light red for the quadrupole and brown for the octopole. Background colours also indicate if any of the multipoles over-lap; yellow for no overlapped multipoles (right column and bottom of middle column), green for overlapped dipole and octopole, light yellow (bottom of left column) for quadrupole and octopole and light blue (top of left column) if all the multipole vectors are overlapped.}\label{MvsMaps}
\end{centering}
\end{figure}

Figure \ref{MvsVz0} shows the multipole vectors from the Bianchi V map which serves as a good example to show how strongly the multipoles are correlated. The dipole (left) lies exactly at the top of the sphere, which is at the centre of the image. The two quadrupole vectors (middle) are located on the same spots on which two of octopole vectors (right) are placed. The remaining octopole vector is in the centre, i.e. the same place as dipole. Now we plot
the dipole, quadrupole and octopole on the same 2-sphere for all the Bianchi maps, at different redshifts, to see how exactly they overlap (see Figure \ref{MvsMaps}). The background colours indicate if any of the multipoles over-lap; yellow for no overlapped multipoles, green for overlapped dipole and octopole, light yellow for overlapped quadrupole and octopole, and light blue if all the multipole vectors are overlapped. The z-axis is into the page and
the x-y plane is the large marked circle.

First of all, in all types of the Bianchi models, we see the
quadrupole and octopole vectors lie on the same plane, except for
one of the octopole vectors which is always located in the centre of
the image. For the Bianchi V and VII$_h$ types, the dipole vectors
lie very near the centre in the early stage but not exactly on it.
However, as time goes on, the dipole vector is overlapped by one of
the octopole vectors in the centre. The dipole vector of Bianchi
VII$_0$ type is different from Bianchi VII$_h$. In the Bianchi
VII$_0$ type, there is no particular correlation between the dipole
and other multipoles since the dipole is not coupled with the other
multipoles (quadrupole and octopole). For the Bianchi V (left), at
the beginning of the stage, the dipole vector is `almost' on the
z-axis and one of octopole vectors is exactly on the z-axis. However
the two later-time Bianchi V cases on the top left of Figure
\ref{MvsMaps} show the dipole and octopole are on the same spot, in
the middle of the image, which is exactly on the z-axis. Meanwhile,
one of components of quadrupole and octopole vectors are on the same
spot on the x-y plan. This means that the later Bianchi V models
have more extreme correlation between the multipoles.

\subsection{Results for WMAP Data}
\begin{figure}
\begin{centering}
  \includegraphics[width=41mm]{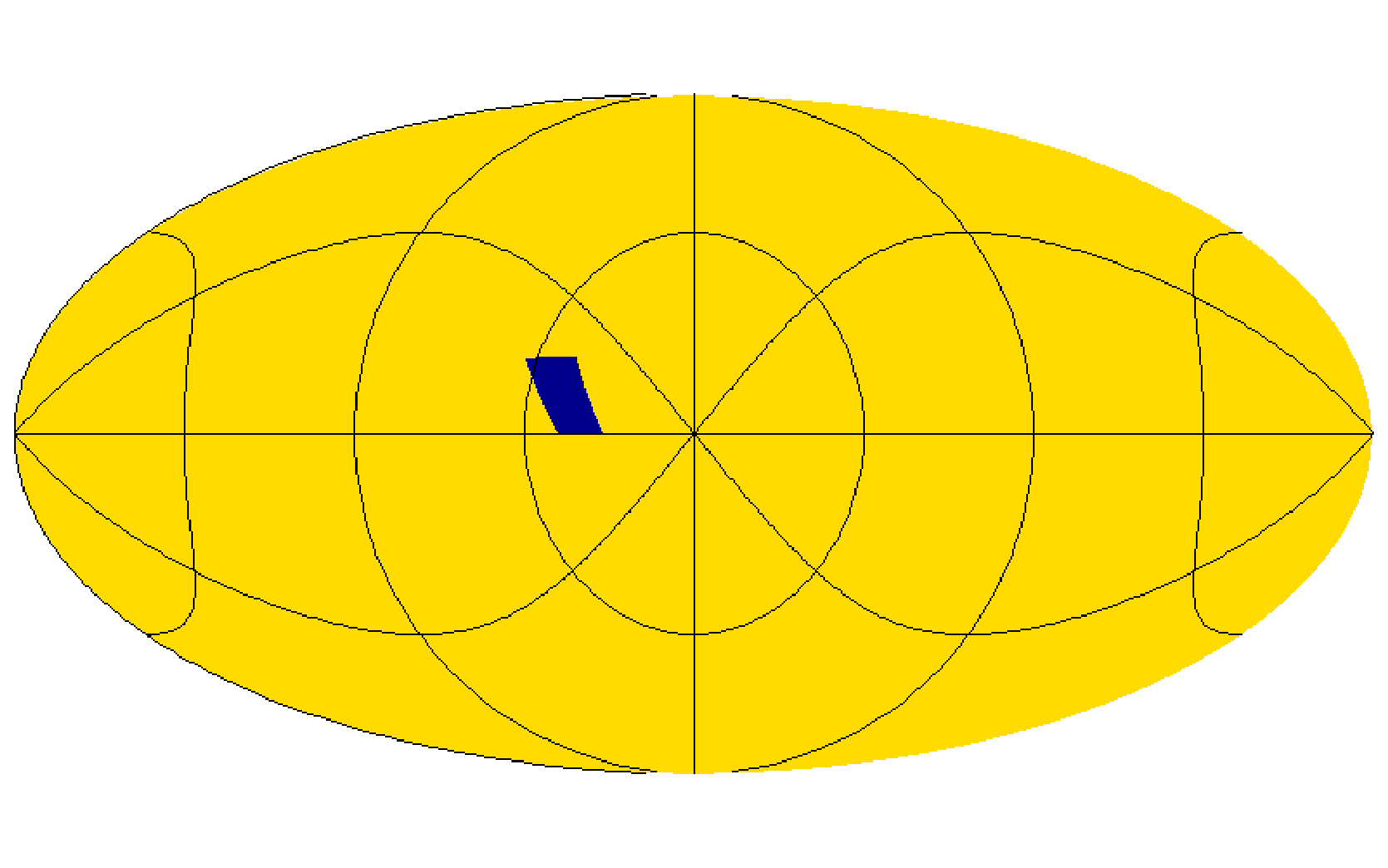}
  \includegraphics[width=41mm]{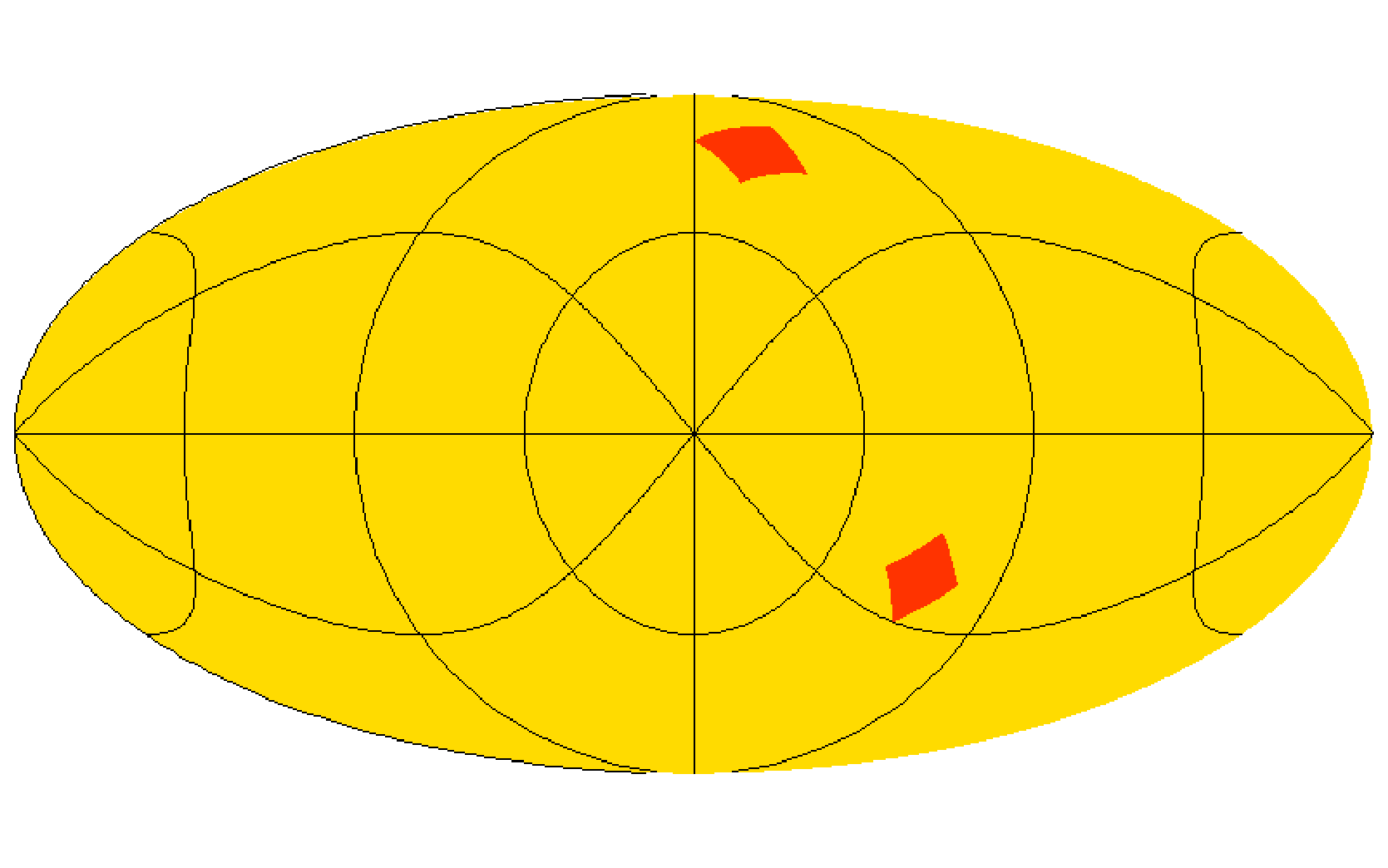}
  \includegraphics[width=41mm]{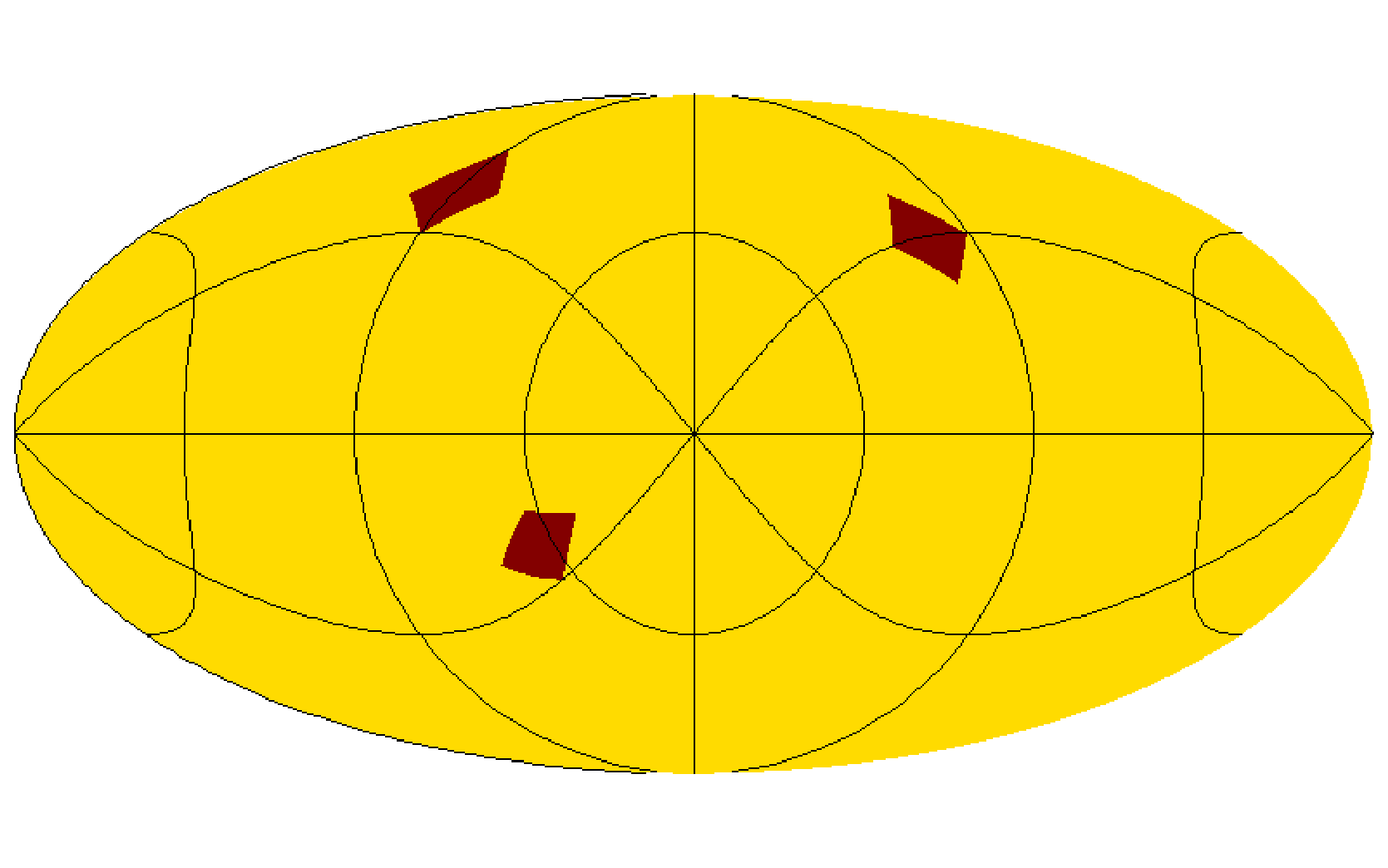}
\\\includegraphics[width=41mm]{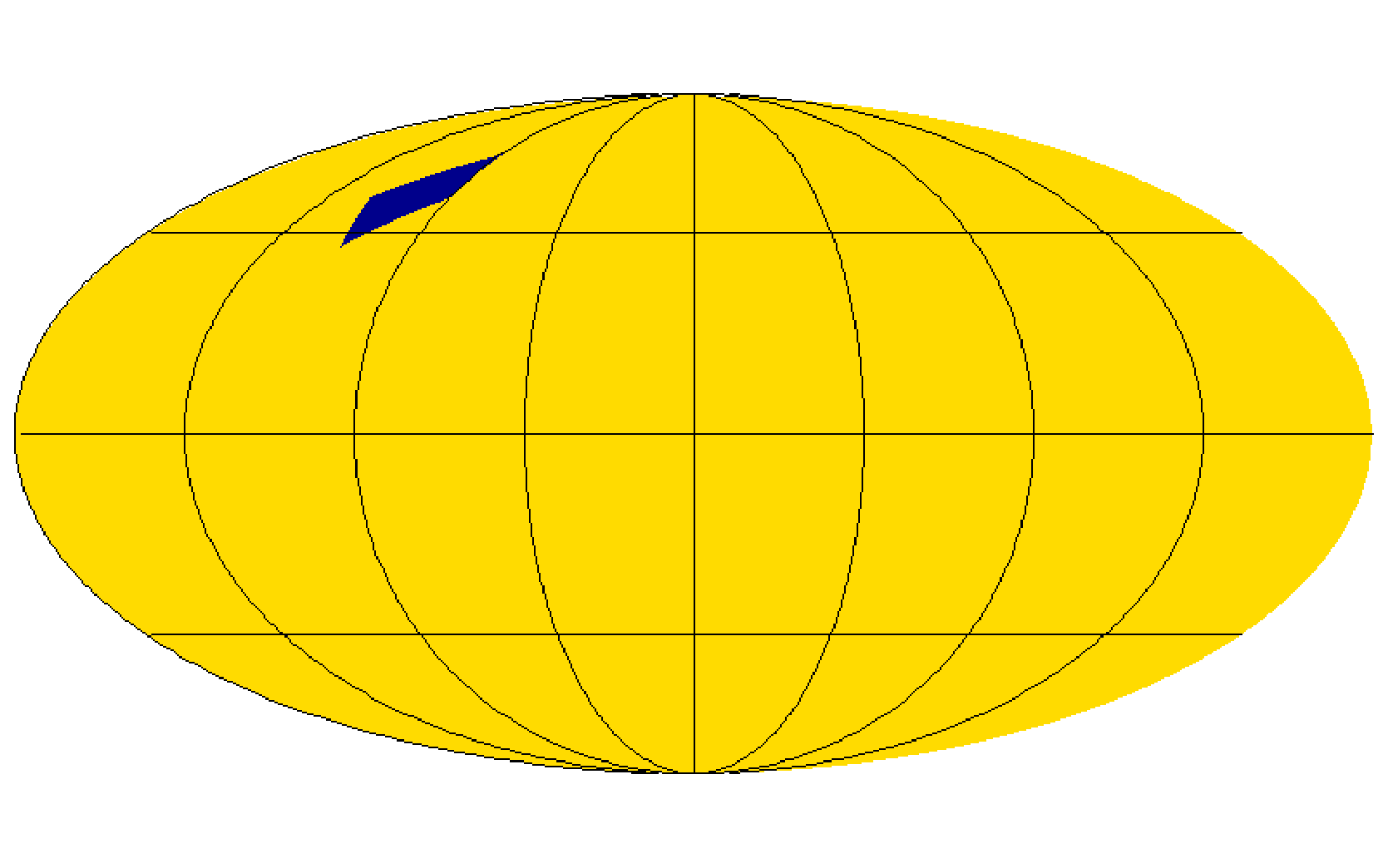}
  \includegraphics[width=41mm]{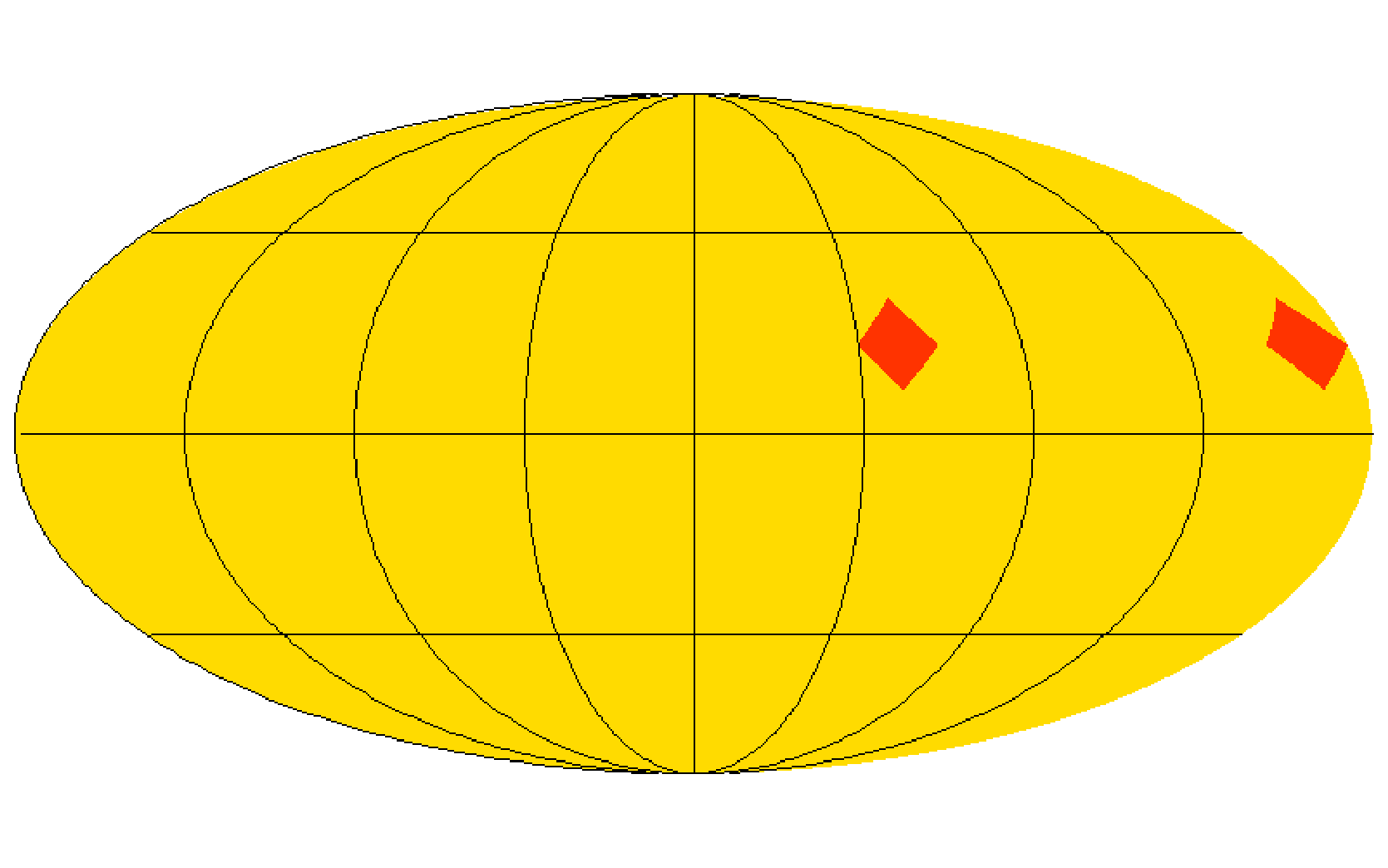}
  \includegraphics[width=41mm]{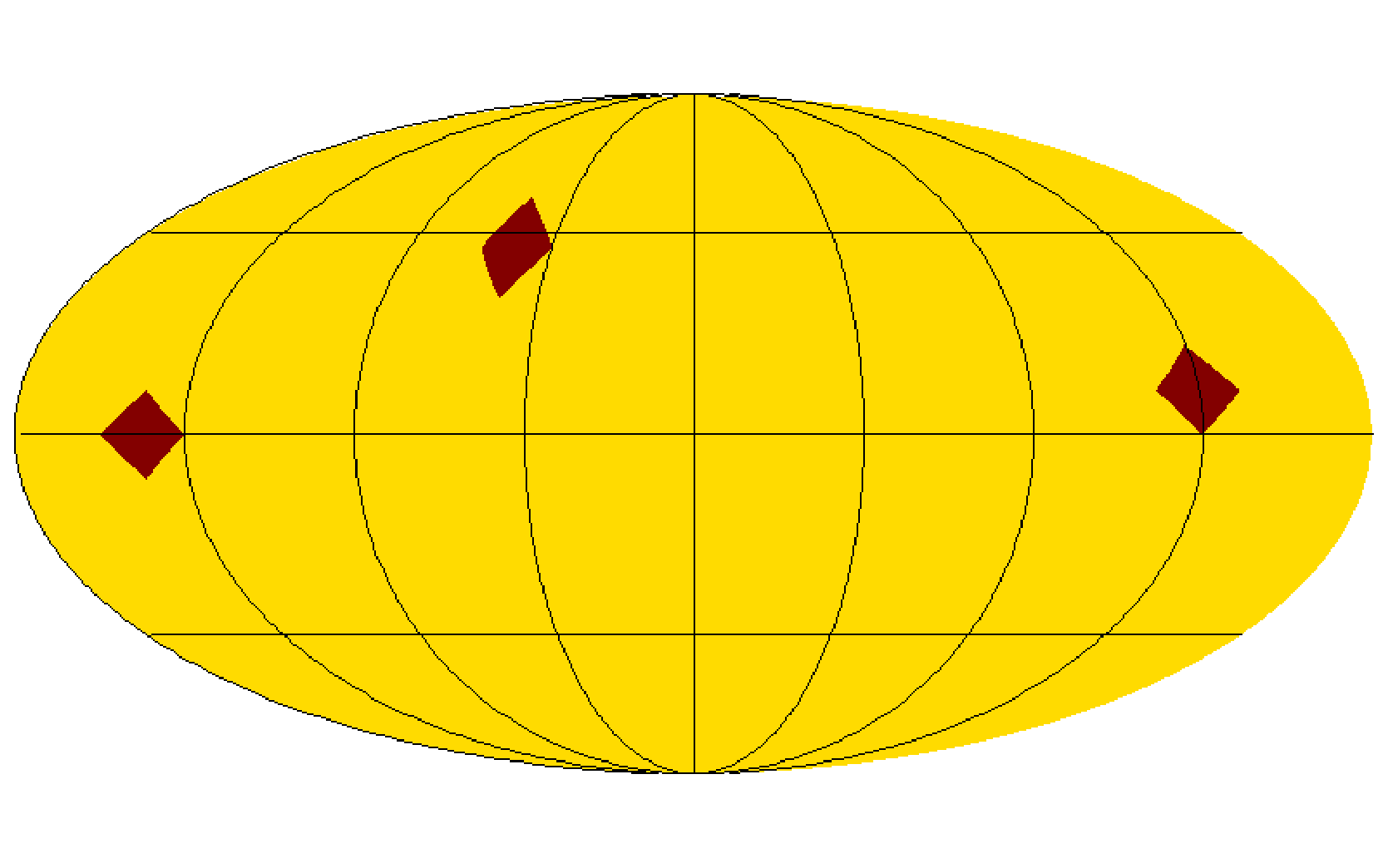}
  \caption{The multipole vectors from the WMAP 5 year ILC map: dipole (left), quadrupole (middle) and octopole (right). The top row shows results where the z-axis is into the page and the x-y plane is the large marked circle. The bottom row shows results where the z-axis is the vertical line and the x-y plane is the horizontal line across the centre of the map.}
  \label{figmvWMAP}
\end{centering}
\end{figure}

Again for pedagogical interest, the multipole vectors are applied to the WMAP 5 year Internal Linear Combination (ILC) map in Figure \ref{figmvWMAP}. At first glance, the results in the top row look like they might be clustered in a similar way to those in the Bianchi models (see Figure \ref{MvsMaps}). However when we look at the results in the second row, where the map is orientated in the usual galactic coordinates, we see the multipoles line up along the x-y plane. This suggests an alternative explanation that the clustering near the x-y plane could be a feature of the residual galactic contamination which is known to remain in the full sky ILC maps. 

\section{Discussion and Conclusions}
\label{secConclu} The aim of this article was to explore some
simple ways of characterizing the large-scale temperature patterns
in CMB maps generated in anisotropic Bianchi type V, VII$_h$ and
VII$_0$ universes. The ultimate purpose of investigating this
behavior is to find ways of quantifying the global properties of the
pattern produced in order to isolate the effect of anisotropy from
that of non-Gaussianity. We repeat that when we talk about
non-Gaussianity here is not related to a stochastic field; there are
no fluctuations in the Bianchi maps. 

We first discussed perhaps the simplest and perhaps the most obvious
possible  descriptive statistics, the histogram of the pixel values,
primarily with the aim of demonstrating how non-Gaussianity of a
sort can arise from asymmetry. We evaluated the pixel distribution
functions for each of the maps and compared them to results expected
in a universe consistent with the concordance model. The type
VII$_0$ maps show the strongest deviation from the null hypothesis;
but types V and VII$_h$ behaved in a similar fashion to each other,
and closer to that of the null hypothesis. The reason these two gave
lesser indications of the presence of anomalies was because the
focussing effect produces a pattern that covers only a smaller part
of the celestial sphere, which tends to get lost when averaged over
the whole sky. This method is therefore useful to characterize
coherent signals extended over a large region, such as a spiral
pattern, but not if they are concentrated.

Phase analysis is a relatively new technique, and has consequently
not been used to quantify many alternative situations to the
concordance model. The phases of the spherical harmonic coefficients
provide a generic way of looking at correlations in harmonic space
that could arise from non-stationarity or non-Gaussianity. While
this is a potential strength of the approach - while phase
correlations will not just be useful for identifying anisotropies
specific to the Bianchi models, but in theory any isotropy
introduced to the CMB - it could also prove a weakness, in that more
general methods may lack the power to discriminate very specific
models.

The phase correlations identified in our Bianchi maps using this
technique were much stronger than we at first expected; given the
generic nature of the metric it was not expected to yield good
results. In addition to this, the strong correlations were found to
be robust to both rotation and moderate noise. Significant correlations in
both twisted and focusing features were also identified. However
using the same methods on the WMAP 5 year data shows little evidence
of non-Gaussianity. Given that the diagnostics are identified in
harmonic space, it is difficult to say whether any of the anomalies
identified this way are down to isotropy or homogeneity.

The analysis of multipole vectors is also a relatively new
technique.  It has been used to identify non-Gaussianities in the
WMAP data, and has been particularly successful in identifying
anisotropies (i.e. asymmetries and/or preferred directions). The
multipole vectors are calculated from spherical harmonic
coefficients which, as we have already shown, themselves provide a
very effective way of identifying correlations in Bianchi (and
presumably other anisotropic) patterns. The multipole vectors must
include at least some of the information needed to describe these
mode correlations. The advantage of multipole vectors over the
spherical harmonics themselves, however, is that they give results
in real (i.e. pixel) space which is much more informative to the
user. The results when applied to the Bianchi maps show very strong
correlations between the directions of the multipole vectors for low
$\l$, often with them entirely overlapping, and hence showing
preferred directions. Since these vectors would not be aligned in
the case of a stationary stochastic field over the sky, these
results demonstrate that they are sensitive to departures from the
standard cosmological model.

It remains the case that the standard cosmological model is a  good
fit to a huge range of observational data. Nevertheless, it is
important that tools are developed that are sufficiently sensitive
to hunt efficiently for possible anomalies in the next generation of
observations. There are many ways that the CMB temperature pattern
could be anomalous other than through the presence of Bianchi
perturbation modes such as those we have studied here. Just as there
are many ways a distribution can be non-Gaussian, so are there also
many ways a fluctuation field can be non-stationary. Testing for
departures from the standard model will require not one but a
battery of statistical techniques each sensitive to particular
aspects of the distribution.

This has been a very preliminary analysis, aimed at  establishing
whether the diagnostics described in this paper are {\em in
principle} capable of uncovering evidence of underlying anomalies in
CMB data. Of course these patterns represent somewhat extreme
departures from the standard framework so it is no real surprise
that they register strongly in the descriptors used. However, in all
cases our analysis has involved only a relatively small number of
quantities, so the fact that we see quantifiable effects emerging is
very encouraging.

\section*{Acknowledgments}
We acknowledge the use of the  Legacy Archive for Microwave
Background Data Analysis (L\textsc{ambda}) and many of the
calculations in this paper made use of the \healpix package
\citep{heal}. Jo Short receives funding from an STFC studentship.
Rockhee Sung acknowledges an Overseas Scholarship from the Korean
government.

\appendix
\section{Multipole vectors terminology}
\label{appMV}
Any homogeneous polynomial $F_R$ of degree $\l$ in $x$, $y$ and $z$
can be written as
\begin{equation}
F_R(x,y,z) = \lambda\cdot(a_1x+b_1y+c_1z)\cdot(a_2x+b_2y+c_2z)
\cdots (a_\l x+b_\l y+c_\l z) + S_R\cdot G_R.
\end{equation}
The function $G_R$ is homogeneous of degree $\l-2$ and
$S_R=x^2+y^2+z^2$. All polynomials $F_R$, $G_R$ and $S_R$ and all
variables are real. If we consider the values taken by the
polynomial on the unit sphere where $S_R=x^2+y^2+z^2=1$ then the
above expression reduces to the product of the linear parts together
with the term $G_R$. This can be extended to a complex polynomial
through
\begin{equation}\label{FRMVS}
F(x,y,z) =\lambda L_1L_2\cdots L_\l+S\cdot G
\end{equation}
where $L_i=a_ix+b_iy+c_iz$ and $x$, $y$ and $z$ are complex numbers,
while the coefficients of $F$ ($a_i,b_i,c_i,$ and $ \lambda$) are
real. $G$ is a homogeneous polynomial of degree $\l-2$ and $S$ is
given by $x^2+y^2+z^2$. We are interested in the value of the
polynomial $F$ on the 2-sphere which, in the complex space, takes
the form $S=x^2+y^2+z^2=0$. By B\'{e}zout's theorem, which holds
that the number of points on two curves is equal to the product of
their degrees, there are 2$\l$ points in which the complex curve
$F(x,y,z)=0$ intersects the quadratic curve $S(x,y,z)=0$, which is
topologically a 2-sphere. Since the complex curve $F=0$ intersects
the complex $S=0$ in 2$\l$ points, the product curves $L_i$ also
intersect $S=0$ in the same 2$\l$ points, $f_i=(x_i,y_i,z_i)$.
Moreover, its complex conjugate $f^*_i=(x^*_i,y^*_i,z^*_i)$ has to
lie in the intersection in which both $F$ and $S$ are zero. We thus
obtain the 2$\l$ points of intersection $\{
f_1,f^*_1,\cdots,f_\l,f^*_\l\} $.

Each pair $\{f_\l,f^*_\l\}$ determines a unique line
$L_i=a_ix+b_iy+c_iz=0$ with real coefficients such that:
\begin{eqnarray}\label{fi0}
a_ix^{Re}_i+b_iy^{Re}_i+c_iz^{Re}_i &=&0 \nonumber\\
a_ix^{Im}_i+b_iy^{Im}_i+c_iz^{Im}_i &=&0
\end{eqnarray}
where the coefficients are normalized to unit length, i.e.
$a^2_i+b^2_i+c^2_i=1$. Note that the index $i$ has no sum and $1\leq
i \leq \l $. In order to find the multipole vectors, $v_i=(
a_i,b_i,c_i )$, of each $\l$ we need to find the pairs of $f_i$
which lie on the curve $F=0$ and $S=0$ in the complex projective
plane i.e. finding the roots ($\alpha$) which satisfy $F=0$ on the
2-sphere ($S=0$).  What is required is to factorize the homogeneous,
harmonic polynomial, $F$ into linear factors i.e. such that $F$ is
the product of $L_i$ only. However, this is not possible
analytically since they are not linear equations far from the
dipole. Fortunately, the curve $S=0$ can be parametrized as a
single variable and the polynomial $F$ as:
\begin{equation}\label{Fc0}
F(x,y,z)=F(i(\alpha^2-1),-2i\alpha,\alpha^2+1).
\end{equation}
From Equation \ref{Fc0}, the roots $\alpha$ which satisfy $F=0$ and
$S=0$, or the product of $L_i$=0, can be found. Once the roots
$\alpha$ are found, the pair of $\{f_i,f^*_i\}$ can be expressed in
terms of x, y and z. The next step is to find the multipole vectors
from Equation \ref{fi0} by using $f_i=(x^{Re}, y^{Re},
z^{Re})+i(x^{Im}, y^{Im}, z^{Im})$ or its conjugate $f^*_i$ since
the two points give same result.

We now apply this terminology to the relevant cosmological
application, that of temperature patterns on the CMB sky, as
follows. The $\l^{th}$ multipoles, $T_\l$, can be represented by a
polynomial $F$ on the 2-sphere ($S=0$), or the product of $L_i$. By
the given relations $x=i(\alpha^2-1)$, $y=-2i\alpha$, and $
z=\alpha^2+1$, the spherical harmonics can be expanded as $\alpha$
terms. Thus the dipole ($T_1$), quadrupole ($T_2$) and octopole
($T_3$) can be described in terms of $\alpha$  with the $\alm$ as
coefficients. The $\alm$ were calculated using \healpix from the
maps, and we found the roots $\alpha$ which satisfy equations
$T_\l$=0 for each $\l$. These roots gave a pair of $\{f_\l,f^*_\l\}$
, therefore from Equation \ref{fi0} we obtained solution sets
$v_i=(a_i,b_i,c_i)$ which are multipole vectors for each $\l$.

We now explain this procedure in detail for each of the multipoles
considered:
\subsubsection{Dipole: $\l=1$.}
\begin{eqnarray}
T_1&=& \sum_m a_{1m}Y_{1m} \\
   &=& \sqrt{\frac{3}{2\pi}}(a^{Re}_{11}x-a^{Im}_{11}y+\frac{a_{10}}{\sqrt{2}}z )
\end{eqnarray}
For the dipole vector it is not necessary to use the $\alpha$ notation
since it is a linear equation in $x$, $y$ and $z$. In our polynomial
notation, the dipole is also represented as
\begin{eqnarray}
F=L_1=\lambda_1(a_1x+b_1y+c_1z).
\end{eqnarray}
From the normalization, we obtain the multipole vector,
\begin{equation}
v_1=
(\sqrt{\frac{2}{3C_1}}a^{Re}_{11},-\sqrt{\frac{2}{3C_1}}a^{Im}_{11},\frac{1}{\sqrt{3C_1}}a_{10}
),
\end{equation}
and the coefficient,
\begin{equation}
\lambda_1=\sqrt{\frac{3}{2\pi}}\sqrt{a^{2Re}_{11}+a^{2Im}_{11}+\frac{1}{2}a^2_{10}}
      =\sqrt{\frac{3}{2\pi}}\sqrt{a^{2}_{11}+\frac{1}{2}a^2_{10}}=\frac{3}{2}\sqrt{\frac{C_1}{\pi}}
\end{equation}
where the C$_1=( 2|a_{11}|^2+a^2_{10} )/3$ is the angular power spectrum C$_\l$ of the monopole.

\subsubsection{Quadrupole: $\l=2$.}

In this case we need to expand $T_2$ using the $\alpha$ notation
since for multipoles from the quadrupole to higher order the
equations are no longer linear. The quadrupole on the sphere has two
multipole vectors, $\upsilon_1$ and $\upsilon_2$, from
$F=L_1L_2=\lambda_1(a_1x+b_1y+c_1z)(a_2x+b_2y+c_2z)$. In order to
find them, we transfer the quadrupole expression from spherical
harmonics to $\alpha$ notation for efficient computing,
\begin{eqnarray}
T_2&=& \sum_m a_{2m}Y_{2m} \\
   &=&( \sqrt{\frac{3}{2}}a_{20}-a^{Re}_{22}+2ia^{Re}_{21})\alpha^4-4(a^{Im}_{22}-ia^{Im}_{21})\alpha^3
      +(\sqrt{6}a_{20}+6a^{Re}_{22} )\alpha^2  \nonumber\\
      && \quad +4(a^{Im}_{22}+ia^{Im}_{21})\alpha +\sqrt{\frac{3}{2}}a_{20}-a^{Re}_{22}-2ia^{Re}_{21}
\end{eqnarray}
\subsubsection{Octopole: $\l=3$.}
We use the same method as we have done for the quadrupole. The
octopole has three multipole vectors, $\upsilon_1$, $\upsilon_2$  and
$\upsilon_3$, from
$F=L_1L_2L_3=\lambda_1(a_1x+b_1y+c_1z)(a_2x+b_2y+c_2z)(a_3x+b_3y+c_3z)$,
which gives a $6^{th}$ order of equation in $\alpha$:
\begin{eqnarray}
T_3&=& \sum_m a_{3m}Y_{3m} \\
   &=& A_6 \alpha^6 +A_5 \alpha^5 +A_4 \alpha^4 +A_3 \alpha^3 +A_2 \alpha^2 +A_1 \alpha
   +A_0,
\end{eqnarray}
in which the coefficients are,
\begin{eqnarray*}
 A_6 &=&5a_{30}-\sqrt{30}a^{Re}_{32}+(-5\sqrt{3}a^{Re}_{31}+\sqrt{5}a^{Re}_{33})i   \\
 A_5 &=& -4\sqrt{30}a^{Im}_{32}+(-10\sqrt{3}a^{Im}_{31}+6\sqrt{5}a^{Im}_{33})i ]\\
 A_4 &=& 5[3a_{30}+\sqrt{30}a^{Re}_{32}-(\sqrt{3}a^{Re}_{31}+3\sqrt{5}a^{Re}_{33} )i \\
 A_3 &=& -20(\sqrt{3}a^{Im}_{31}+\sqrt{5}a^{Im}_{33})i\\
 A_2 &=&A_4^*, A_1 =-A_5^*, A_0 = A_6^*.
 \end{eqnarray*}
Each multipole has 2$\l$ roots of $\alpha$ which gives the $(f_i,
f^*_i)$ pairs, however only $\l$ components are used to find the
multipoles since their conjugators give the same results, as we
mentioned earlier.

\label{lastpage}


\begin{thebibliography}{99}


\bibitem[\protect\citeauthoryear{Abramo et al.}{2006}]{Abramo2006} Abramo L. R., Bernui A., Ferreira I. S., Villela T., Wuensche C. A., 2006, Phys. Rev. D, 74, 063506

\bibitem[\protect\citeauthoryear{Bardeen, Steinhardt \& Turner}{1983}]{Bardeen1983} Bardeen J. M., Steinhardt P. J., Turner M. S., 1983, Phys. Rev. D, 28, 679

\bibitem[\protect\citeauthoryear{Barrow, Juszkiewicz \& Sonoda}{1985}]{Barrow1985} Barrow J. D., Juszkiewicz R., Sonoda D. H., 1985, MNRAS, 213, 917

\bibitem[\protect\citeauthoryear{Bennett et al.}{2003}]{Bennett2003} Bennett C. L.  et al., 2003, ApJS, 148, 1

\bibitem[\protect\citeauthoryear{Bielewicz et al.}{2005}]{Bielewicz2005} Bielewicz P., Eriksen H. K., Banday A. J., G\'{o}rski K. M., Lilje P. B., 2005, ApJ, 635, 750

\bibitem[\protect\citeauthoryear{Bielewicz \& Riazuelo}{2009}]{Bielewicz2009} Bielewicz P., Riazuelo A., 2009, MNRAS, 396, 609

\bibitem[\protect\citeauthoryear{Bridges et al.}{2008}]{McEwen2} Bridges M., McEwen J.D., Cruz M.,
Hobson M.P., Lasenby A.N., Vielva P., Mart\'{i}nez-Gonzalez E.,
2008, MNRAS, 390, 1372

\bibitem[\protect\citeauthoryear{Bunn, Ferreira \& Silk}{1996}]{Bunn1996} Bunn E. F., Ferreira P., Silk J., 1996, Phys. Rev. Lett., 635, 750

\bibitem[\protect\citeauthoryear{Chiang, Naselsky \& Coles}{2004}]{Chiang2004} Chiang L.-Y., Naselsky P. D., Coles P., 2004, ApJ, 602,L1

\bibitem[\protect\citeauthoryear{Chiang et al.}{2007}]{ccno7} Chiang L.-Y., Coles P., Naselsky P. D., Olesen P.,
2007, JCAP, 1, 21

\bibitem[\protect\citeauthoryear{Chiang, Naselsky \& Coles}{2007}]{Chiang2007} Chiang L.-Y., Naselsky P. D., Coles P., 2007, ApJ, 664, 8

\bibitem[\protect\citeauthoryear{Coles et al.}{2004}]{Coles2004} Coles P., Dineen P., Earl J., Wright D., 2004, MNRAS, 350, 989

\bibitem[\protect\citeauthoryear{Coles \& Chiang}{2000}]{cc2000} Coles P., Chiang L.-Y., 2000, Nat, 406,
376

\bibitem[\protect\citeauthoryear{Copi, Huterer \& Starkman}{2004}]{Copi2004} Copi C. J. , Huterer D., Starkman G. D., 2004, Phys. Rev. D, 70, 043515

\bibitem[\protect\citeauthoryear{Copi et al.}{2006}]{Copi2006} Copi C. J. , Huterer D., Schwarz D. J., Starkman  G. D., 2006, MNRAS, 367, 79

\bibitem[\protect\citeauthoryear{Copi et al.}{2007}]{Copi2007} Copi C. J. , Huterer D., Schwarz D. J., Starkman  G. D., 2007, Phys. Rev. D., 75, 023507

\bibitem[\protect\citeauthoryear{Cruz et al.}{2005}]{Cruz2005} Cruz M., Mart\'{i}nez-Gonz\'{a}lez E., Vielva P., Cay\'{o}n L., 2005, MNRAS, 356, 29

\bibitem[\protect\citeauthoryear{Dineen, Rocha  \& Coles}{2005}]{dc2005} Dineen P., Rocha G., Coles P., 2005, MNRAS,
358, 1285

\bibitem[\protect\citeauthoryear{Ellis \& MacCallum}{1969}]{Ellis1969} Ellis G. F. R., MacCallum M. A. H., 1969, Commun. Math. Phys., 12, 108

\bibitem[\protect\citeauthoryear{Eriksen et al.}{2004}]{Eriksen2004a} Eriksen H. K., Hansen F. K., Banday A. J., G\'{o}rski K. M., Lilje P. B., 2004, ApJ, 605, 14

\bibitem[\protect\citeauthoryear{Eriksen et al.}{2005}]{Eriksen2005} Eriksen H. K.  et al., 2005, astro-ph/0508196

\bibitem[\protect\citeauthoryear{Eriksen et al.}{2007}]{Eriksen2007} Eriksen H. K., Banday A. J., G\'{o}rski K. M., Hansen F. K., Lilje P. B., 2007, ApJ, 660, L81

\bibitem[\protect\citeauthoryear{G\'{o}rski et al.}{2005}]{heal} G\'{o}rski K. M., Hivon E., Banday A. J., Wandelt B. D., Hansen F. K., Reinecke M., Bartelmann M., 2005, ApJ, 622, 759

\bibitem[\protect\citeauthoryear{Guth \& Pi}{1982}]{Guth1982} Guth A. H., Pi S. Y., 1982, Phys. Rev. Lett., 49, 1110

\bibitem[\protect\citeauthoryear{Hansen et al.}{2009}]{Hansen2009} Hansen F. K., Banday A. J., G\'{o}rski K. M., Eriksen H. K., Lilje P. B., 2009, ApJ, 704, 1448

\bibitem[\protect\citeauthoryear{Hinshaw et al.}{2009}]{Hinshaw2009} Hinshaw G. et al., 2009, ApJS, 180, 225

\bibitem[\protect\citeauthoryear{Hoftuft et al.}{2009}]{Hoftuft2009} Hoftuft J., Eriksen H. K., Banday A. J., G\'{o}rski K. M., Hansen F. K., Lilje P. B., 2009, ApJS, 699, 985

\bibitem[\protect\citeauthoryear{Jaffe et al.}{2005}]{Jaffe2005} Jaffe T. R., Banday A. J., Eriksen H. K., G\'{o}rski K. M., Hansen F. K., 2005, ApJ, 629, L1

\bibitem[\protect\citeauthoryear{Jaffe et al.}{2006a}]{Jaffe2006a} Jaffe T. R., Banday A. J., Eriksen H. K., G\'{o}rski K. M., Hansen F. K., 2006, ApJ, 643, 616

\bibitem[\protect\citeauthoryear{Jaffe et al.}{2006b}]{Jaffe2006b} Jaffe T. R., Banday A. J., Eriksen H. K., G\'{o}rski K. M., Hansen F. K., 2006, A\&A, 640, 393

\bibitem[\protect\citeauthoryear{Katz \& Weeks}{2004}]{Katz2004} Katz G., Weeks J., 2004, Phys. Rev. D, 70, 063527

\bibitem[\protect\citeauthoryear{Kogut, Hinshaw \& Banday}{1997}]{Kogut1997} Kogut A., Hinshaw G., Banday A. J., 1997, Phys. Rev. D, 55, 1901

\bibitem[\protect\citeauthoryear{Land \& Magueijo}{2005a}]{Land2005a} Land K., Magueijo J., 2005a, MNRAS, 357, 994

\bibitem[\protect\citeauthoryear{Land \& Magueijo}{2005b}]{Land2005b} Land K., Magueijo J., 2005b, MNRAS, 362, L16

\bibitem[\protect\citeauthoryear{Land \& Magueijo}{2005c}]{Land2005c} Land K., Magueijo J., 2005c, Phys. Rev. D, 72, 101302(R)

\bibitem[\protect\citeauthoryear{Land \& Magueijo}{2005d}]{Land2005d} Land K., Magueijo J., 2005d, Phys. Rev. Lett., 95, 071301

\bibitem[\protect\citeauthoryear{Land \& Magueijo}{2005e}]{Land2005e} Land K., Magueijo J., 2005e, MNRAS, 362, 838

\bibitem[\protect\citeauthoryear{Land \& Magueijo}{2007}]{Land2007} Land K., Magueijo J., 2007, MNRAS, 378, 153

\bibitem[\protect\citeauthoryear{McEwen et al.}{2006}]{McEwen1}  McEwen J. D., Hobson M. P., Lasenby A. N., Mortlock D.J.,
2006, MNRAS, 369, 1858

\bibitem[\protect\citeauthoryear{Maxwell}{1891}]{Maxwell1891} Maxwell J. C., A Treatise on Electricity and Magnetism (Clarendon Press, London, 1891), Vol. I, 3rd edition.

\bibitem[\protect\citeauthoryear{Park}{2004}]{Park2004} Park C., 2004, MNRAS, 349, 313

\bibitem[\protect\citeauthoryear{Pontzen}{2009}]{Pontz2} Pontzen A.,
2009, Phys. Rev. D., 79, 103518

\bibitem[\protect\citeauthoryear{Pontzen \& Challinor}{2007}]{Pontz1}
Pontzen A., Challinor A.,  2007, MNRAS, 380, 1387

\bibitem[\protect\citeauthoryear{Schwarz et al.}{2004}]{Schwarz2004} Schwarz D. J., Starkman G. D., Huterer D., Copi C. J., 2004, Phys. Rev. Lett., 93, 221301

\bibitem[\protect\citeauthoryear{Stannard \& Coles}{2005}]{sc2005} Stannard A., Coles P., 2005, MNRAS, 364,
929

\bibitem[\protect\citeauthoryear{Starobinskij}{1982}]{Starobinskij1982} Starobinskij A. A., 1982, Phys. Lett. B, 117, 175

\bibitem[\protect\citeauthoryear{Sung \& Coles}{2009}]{Sung2009} Sung R., Coles P., 2009, CQGra, 26, 172001

\bibitem[\protect\citeauthoryear{Sung \& Coles}{2010}]{Sung2010} Sung R., Coles P., 2010,
JCAP, submitted, arXiv:1004.0957 (astro-ph.CO)

\bibitem[\protect\citeauthoryear{Vielva et al.}{2004}]{Vielva2004} Vielva P., Martínez-González E., Barreiro R. B., Sanz J. L., Cayón L., 2004, ApJ, 609, 22

\bibitem[\protect\citeauthoryear{Yadav \& Wandelt}{2008}]{Yadav2008} Yadav A. P. S., Wandelt B. D., 2008, Phys. Rev. Lett., 100, 181301

\end{thebibliography}
\end{document}